\documentclass[12pt]{article}

\usepackage{setspace}
\usepackage[vmargin = 1.5in, hmargin = 1.5in]{geometry}

\usepackage[usenames,dvipsnames,svgnames]{xcolor}
\usepackage{amsmath, amssymb, amsfonts, graphicx, tikz,pdflscape, mathtools, amsthm,  bm,pgfplots,multirow,  multicol, booktabs, todonotes}
\usepackage[inline]{enumitem}
\usepackage{verbatim}
\usepackage{natbib}

\usepackage{accents}
\newcommand{\ubar}[1]{\underaccent{\bar}{#1}}

\usepackage{needspace}

\definecolor{GmailBlue}{RGB}{42, 93, 176} 
\usepackage[
pagebackref,
colorlinks=true,
citecolor= GmailBlue,
linkcolor=GmailBlue,
urlcolor = GmailBlue
]{hyperref}
\usepackage[capitalize,noabbrev]{cleveref}

\newcommand{\bibtexorder}[1]{}

\usetikzlibrary{calc}
\usepackage[size = small, skip = 10pt]{caption}
\allowdisplaybreaks[1]

\allowdisplaybreaks[1]

\tikzstyle{hollow}=[circle,draw,inner sep=1.5]
\tikzstyle{solid}=[circle,draw,inner sep=1.5,fill=black]

\pgfplotsset{compat = newest}

\usepackage{subcaption}

\def\a{\alpha}
\def\b{\beta}

\def\d{\delta}
\def\e{\varepsilon}

\def\th{\theta}

\def\k{\kappa}

\def\D{\Delta}

\def\R{\mathbf{R}}

\def\KK{\mathcal{K}}

\def\pd{\partial}

\DeclareMathOperator{\E}{\mathbf{E}}
\DeclareMathOperator*{\argmax}{argmax}
\DeclareMathOperator*{\maz}{maximize}

\newcommand{\Paren}[1]{\left( #1 \right)}

\newcommand{\Brac}[1]{\left[ #1 \right]}

\newcommand{\Set}[1]{\left\{ #1 \right\}}

\newcommand{\de}{\mathop{}\!\mathrm{d}}

\newtheoremstyle{break}
{}
{}
{\itshape}
{}
{\bfseries}
{}
{\newline}
{}

\theoremstyle{break}
\newtheorem{thm}{Theorem}
\newtheorem*{theorem*}{Theorem}
\newtheorem*{cor*}{Corollary}
\newtheorem{cor}{Corollary}
\newtheorem{prop}{Proposition}
\newtheorem{lem}{Lemma}

\crefname{prop}{Proposition}{Propositions}
\crefname{thm}{Theorem}{Theorems}
\crefname{lem}{Lemma}{Lemmas}

\theoremstyle{definition}

\newtheorem*{rem*}{Remark}

\usepackage[titletoc]{appendix}

\usepackage{comment}

\usepackage{pgfplots}
\usepgfplotslibrary{groupplots}
\pgfplotsset{compat=newest}
\usepackage{tikz}
\usetikzlibrary{arrows,shapes,trees,patterns,backgrounds,shadows,chains,matrix,calc,3d, mindmap,arrows.meta,positioning}

\date{19 November 2023}
\title{Benefiting from Bias: Delegating to Encourage Information Acquisition\thanks{This paper was previously circulated under the titles ``Biased and Uninformed: Delegating to Encourage Information Acquisition" and ``Benefiting from Bias." For insightful comments and discussions, we are grateful to Kyle Bagwell, Dirk Bergemann, Tilman B\"orgers, Hans Peter Gr\"uner, Marina Halac, Johannes H\"{o}rner, Navin Kartik, Eric Maskin, Volker Nocke, Martin Peitz, Larry Samuelson, Nicolas Schutz, and Thomas Tr\"oger, as well as audiences at Cambridge, Mannheim, and Yale. }}
\author{Ian Ball\thanks{MIT Department of Economics. Email: ianball@mit.edu} \hspace{6mm} Xin Gao\thanks{Department of Economics, University of Birmingham, UK. Email: x.gao.1@bham.ac.uk}}

\providecommand{\keywords}[1]{\textbf{Keywords:} #1}

\begin{document}

\maketitle

\begin{abstract}
A principal delegates decisions to a biased agent. Payoffs depend on a state that the principal cannot observe. Initially, the agent does not observe the state, but he can acquire information about it at a cost. We characterize the principal's optimal delegation set.  This set features a cap on high decisions and a gap around the agent's ex ante favorite decision. It may even induce ex-post Pareto-dominated decisions. Under certain conditions on the cost of information acquisition, we show that the principal prefers delegating to an agent with a small bias than to an unbiased agent. 
\end{abstract}



\newpage
\section{Introduction}

Delegation is ubiquitous. The leader of an organization cannot make every decision, so she must delegate: CEOs delegate to middle managers and politicians delegate to policy advisors. In the classical model of delegation \citep{Holmstrom1977}, a principal delegates decisions to an agent with better information but biased preferences.  The principal faces a tradeoff. To utilize the agent's private information, the principal must give the agent discretion, but discretion allows the agent to bias the decision in his own favor. 


In many applications, the agent does not initially have better information, but he can acquire information at a cost. For example, managers gather information about employee performance, and advisors conduct policy research. In these settings, offering the agent discretion motivates him to acquire information. Conversely, if the agent has little control over the decision, he has little to gain from additional information. 


In our model, an uninformed principal (she) delegates a single-dimensional decision to an agent (he) who is biased toward higher decisions. The agent is \textit{initially} uninformed of the state. After the agent observes the delegation set, he chooses a level of costly private effort on experimentation. This effort determines the probability with which the experiment succeeds. If the experiment succeeds, the agent learns the state. Otherwise, the experiment fails and the agent learns nothing. After the experiment's outcome is realized, the agent chooses a decision from the delegation set. 

We first characterize the form of optimal delegation. \cref{res:characterization} says that any optimal delegation set must take one of three forms: \textit{hollow}, \emph{interval}, and \textit{high-point}. A hollow set has a gap around the agent's ex ante favorite decision and a cap that prohibits high decisions.  The gap lowers the agent's payoff from remaining uninformed, thus motivating the agent to acquire information.  The cap restricts the agent's bias if he learns that the state is high. If the optimal cap is low enough to exclude the agent's ex ante favorite decision, then there is no gap, and the optimal delegation set is an interval. Finally, a high-point set consists of an interval together with a singleton decision (the high point) that the principal and agent agree is too high in every state. The agent chooses the high point only if he learns that the state is very high. Including the high point induces the agent to acquire more information, and this benefit outweighs the direct loss from the agent's higher decisions.


In \cref{res:delegation_regime}, we provide conditions under which the delegation set takes the different forms. Hollow delegation is optimal if the bias is sufficiently small. Interval delegation is optimal if the bias is sufficiently large and the agent's optimal effort choice is sufficiently concave as a function of the return from learning the state.  We show by example that high-point delegation can be optimal if information acquisition is very costly. 

Finally, we analyze the principal's preferences over the agent's level of bias. In the classical delegation problem with an informed agent, the principal wants the agent's bias to be as small as possible. In our setting, the agent's bias can benefit the principal by creating a wedge between the ex ante favorite decisions of the principal and the agent. With this wedge, the principal can select a delegation set that punishes the agent if his experiment fails, at a lower cost to the principal. \cref{res:CS} shows that for a range of cost functions, the principal prefers delegating to an agent with a small bias than to an unbiased agent.

The rest of the paper is organized as follows. \cref{sec:literature} discusses related literature. \cref{sec:model}  presents the model. \cref{sec:optimization} analyzes the players' optimization problems. In \cref{sec:characterization}, we characterize the three forms that optimal delegation can take. We then give conditions under which each form is optimal. \cref{sec:bias} studies the principal's preferences over the agent's level of bias. \cref{sec:conclusion} is the conclusion. Proofs are in \cref{sec:proofs}. 

\subsection{Related literature} \label{sec:literature}

Our paper connects the classical delegation literature with more recent work on information acquisition. \cite{Holmstrom1977} introduces the delegation problem with an informed agent. He proves that interval delegation is optimal in the UQC setting---\textit{uniform} state distribution,  \textit{quadratic} losses, and \emph{constant} (state-independent) bias. Subsequent work shows that interval delegation remains optimal under successive relaxations of the UQC assumptions. In a uniform-quadratic setting with affine bias, \cite{melumad1991communication} characterize the incentive-compatible decision rules. They show that interval delegation is optimal as long as the bias is not very sensitive to the state. If the bias is sufficiently sensitive to the state, then a two-point delegation set is optimal.  In a quadratic setting with constant bias, \cite{martimort2006continuity} give a sufficient condition on the state distribution for interval delegation to be optimal.  In a quadratic setting with arbitrary bias, \cite{alonso2008optimal} characterize whether delegation is valuable. They provide a general characterization of the optimal delegation set,  and they give a condition that is necessary and sufficient for interval delegation to be optimal. In a more general setting that allows for non-quadratic preferences, \cite{amador2013theory} give separate necessary and sufficient conditions for interval delegation to be optimal.

We work in the tractable setting of  \cite{Krahmer2016sequential}; we discuss the contribution of their paper below.  This setting allows for a non-uniform state, non-quadratic losses, and non-constant bias.\footnote{\cite{Kleiner2022} uses a similar decision setting to study multi-dimensional delegation.} In this setting, if the agent were informed, then the optimal delegation set would be an interval.

In our model, the agent is biased and information acquisition is costly. Both features are necessary to identify the principal's benefit from bias. Earlier work studies information acquisition and bias separately. \citet{szalay2005economics} studies delegation to an unbiased agent who can pay a cost to privately learn about the state.\footnote{Similarly, in \cite{DemskiSappington1987}, the agent can acquire information at a cost, but the principal cannot (it would be prohibitively costly). In their model, the principal cannot restrict the delegation set, and transfers are allowed.}  Under certain conditions, the optimal delegation set has a gap around the players' ex ante favorite decision.\footnote{\cite{aghion1997formal} observe that giving the agent discretion can encourage information acquisition, but they do not study a constrained delegation problem.} This gap encourages the agent to acquire information. \cite{semenov2018delegation} studies delegation to a biased agent who, prior to contracting, privately learns the state with some exogenous probability. The agent cannot choose to acquire additional information.\footnote{As an extension, \cite{semenov2018delegation} allows the agent to acquire information at a cost.  The paper does not analyze the design of the delegation set in that setting. The paper suggests that the characterization of optimal delegation continues to hold, but this overlooks the effect of the delegation set on the agent's choice of effort.} The optimal delegation set has a gap around the agent's favorite uninformed decision. If the agent has not learned the state, then this gap induces him to select a less biased decision.


\cite{Krahmer2016sequential} study a delegation problem in which the agent initially observes a private, binary signal of the state. Then, after contracting, the agent learns the state perfectly. Their paper focuses on whether sequential screening---by offering a menu of delegation sets---outperforms the optimal static delegation set. The optimal static delegation set is an interval, but if sequential screening is strictly optimal, the optimal menu contains non-interval delegation sets. In their model, static interval delegation is optimal if the agent's bias is sufficiently small. In our model, hollow delegation is optimal if the agent's bias is sufficiently small. Finally, \citet{Ivanov2010informational} compares delegation with cheap-talk in a setting in which the principal can design the agent's private information.\footnote{In a fixed UQC setting, \cite{Goltsman2009mediation} compare delegation with two other protocols: mediation and negotiation (multi-round cheap talk).}


Beyond the setting of delegation, misaligned preferences can bring benefits for different reasons.\footnote{\cite{li2001theory} and \cite{gerardi20008costly} show that an undesirable default option can encourage information acquisition. In a model of delegating regulations, \cite{bubb2014optimal} study a different form of preference bias that directly increases the agent's marginal benefit from learning the state. In a macroeconomic model, \cite{rogoff1985optimal} shows that appointing a central banker who is biased towards inflation stabilization can mitigate the problem of time-inconsistency.} In communication games with endogenous information acquisition, the sender may acquire more information if he is more biased \citep{argenziano2016strategic} or has more divergent beliefs \citep{che2009opinions}. Our analysis is quite different  because the principal (receiver) has commitment power.  




\section{Model} \label{sec:model}

\subsection{Delegation setting}

There are two players: a principal (she) and an agent (he). The principal controls a decision $y \in \R$.  Payoffs from the decision depend on a state $\th \in \R$, drawn from a commonly known distribution function $F$ with continuous, strictly positive density $f$ on its support $[\ubar{\th}, \bar{\th}]$. The principal and agent have state-dependent utilities $u_P(y, \th)$ and $u_A(y, \th)$, which will be specified in \cref{sec:decision_prefs}.

The principal does not observe the state. Initially, the agent does not observe the state either, but he can privately experiment at a cost. As in \cite{szalay2005economics}, the agent chooses experimentation effort $e \in [0,1)$. This effort level determines the probability with which the experiment succeeds. With probability $e$, independent of the state, the experiment succeeds and the agent learns the state. Otherwise, the experiment fails and the agent does not learn the state. The principal does not observe the agent's effort choice or the realization of the experiment. The rest of the setting is common knowledge. 

The agent has a twice continuously differentiable effort cost function
\[
	 c \colon [0,1) \to \R,
\]
satisfying $c(0) = c'(0) = 0$; $c''(e) > 0$ for all $e$ in $(0,1)$; and $\lim_{e \uparrow 1} c'(e) = \infty$.


\begin{figure}
	\centering
	\begin{tikzpicture}[every text node part/.style={align=center}]
		\draw[-|,very thick] (0,0) --(1,0);
		\draw[-|,very thick] (1,0) --(4,0);
		\draw[-|,very thick] (4,0) --(7,0);
		\draw[very thick, -|] (7,0) --(10,0);
		\draw[->,very thick] (10,0)--(11,0);
		\node at (1,-0.2) [below] {principal selects \\ delegation set $D$}; 
		\node at (4,0.2) [above] {agent chooses\\ effort $e$};
		\node at (7,-0.2) [below] {with prob $e$,\\ agent learns $\th$};
		\node at (10,0.2) [above] {agent chooses \\ decision $y \in D$};
	\end{tikzpicture}
	\caption{Timing}
	\label{fig:timing}
\end{figure}

\cref{fig:timing} shows the timing.  First, the principal selects a compact delegation set $D$.\footnote{Offering a delegation set is equivalent to committing to a deterministic decision rule on an abstract message space.  \cite{KovacMylovanov2009stochastic} give conditions under which stochastic mechanisms cannot improve upon deterministic mechanisms; see \cref{ft:weaker}. For an analysis of delegation with money-burning, see \cite{amador2013theory} and \cite{AmbrusEgorov2017}} The agent observes the set $D$ and chooses effort $e$. Next, the outcome of the experiment is realized. The agent observes whether the experiment succeeds. He observes the state if and only if the experiment succeeds. Then he selects a decision $y$ from $D$, and payoffs are realized.

\subsection{Decision preferences} \label{sec:decision_prefs}

Following \cite{Krahmer2016sequential},\footnote{\label{ft:normalization}\cite{Krahmer2016sequential} put bias $-b ( \th)$ in the principal's utility; we put bias $b(\th)$ in the agent's utility. That is, we normalize the state relative to the \emph{principal's} utility. Our normalization makes some formulas more complicated, but it allows us to vary the agent's bias, without changing the state distribution.} utilities for the principal and agent are given by 
\begin{equation*}
\begin{aligned}
u_P(y,\theta) &=  \th y + a(y),\\
u_A(y,\theta) &=  (\th + b(\th)) y + a(y),
\end{aligned}
\end{equation*}
where $a \colon \R \to \R$ is strictly concave and twice continuously differentiable, and $b \colon [\ubar{\th}, \bar{\th}] \to \R$ is continuously differentiable and satisfies $b'(\th) > - 1$ for all $\th$. Thus, $\th +b(\th)$ is strictly increasing in $\th$. This setting nests quadratic-loss utilities.\footnote{Take $a (y) = - (1/2) y^2$. Add the decision-irrelevant terms $- (1/2) \th^2$ to $u_P$ and $-(1/2) ( \th  + b(\th))^2$ to $u_A$. Then scale these utilities by $2$.}

To ensure that the principal and agent have utility-maximizing decisions in each state, we assume
\[
\lim_{y \to - \infty} - a'(y) <  \ubar{\th}  \wedge (\ubar{\th} + b(\ubar{\th})), 
\qquad
\lim_{y \to \infty} -a'(y) > \bar{\th} \vee (\bar{\th} + b(\bar{\th})),
\]
where $\wedge$ and $\vee$ denote the minimum and maximum operators, respectively. 

In each state $\th$, the principal's and the agent's favorite decisions, denoted $y_P(\th)$ and $y_A(\th)$, are defined by the first-order conditions
\[
	- a'( y_P(\th)) = \th, \qquad - a'(y_A(\th)) = \th + b(\th). 
\]
Their ex ante favorite decisions, denoted $y_{P,0}$ and $y_{A,0}$, are given by 
\[
	- a' ( y_{P,0}) = \E [ \th], 
	\qquad 
	- a'(y_{A,0}) = \E [\th + b(\th)].
\]
The functions $y_A$ and $y_P$ are strictly increasing.  Since $y_A(\th) = y_P (\th + b(\th))$, we interpret $b(\th)$ as the agent's bias in state $\th$. In particular, we have  $y_A (\th) \geq y_P (\th)$ if and only if $b(\th) \geq 0$. 

We maintain from \cite{Krahmer2016sequential} the following joint assumptions on the bias function and state distribution.\footnote{Our expressions look different because of our different normalization; see \cref{ft:normalization}.} Let $B(\th) = b(\th)/ (1 + b'(\th))$. Recall that
$1 + b'(\th) > 0$, so $B(\th)$ is well-defined and has the same sign as $b(\th)$. 
\begin{enumerate}[label = A\arabic*., ref = A\arabic*]
		\item \label{it:f} $f(\th) + (B(\th) f(\th))' > 0$ for all $\th$.
		\item \label{it:B} $b(\ubar{\th}) \geq 0$; $b(\bar{\th}) > 0$; and $\E [b(\th)] > 0$.
		\item \label{it:cap} $\ubar{\th} + b(\ubar{\th}) < \E[\th]$. 
\end{enumerate}

Assumption \ref{it:f} ensures that in the informed-agent delegation problem, interval delegation is optimal. In each state $\th$, the inequality guarantees that if the induced decision rule has a jump at state $\th$, then the principal can strictly increase her payoff by adding the decision $y_A(\th)$ to the delegation set.  In the special case of constant bias $b(\th) = \b > 0$, Assumption~\ref{it:f} reduces to the inequality $(\log f(\th) )' = f'(\th)/f(\th) > - 1/\b$. That is, the density $f$ does not decay (multiplicatively) too quickly. The smaller the agent's bias, the more permissive is this constraint. 

Assumptions \ref{it:B} and \ref{it:cap} are less substantive. Under Assumption \ref{it:f}, the optimal delegation set for an informed agent is an interval. Assumption~\ref{it:B} ensures that this interval excludes high decisions, but not low decisions. Assumption \ref{it:cap} ensures that this interval is not a singleton. With constant bias $b(\th) = \b$, Assumptions \ref{it:B} and \ref{it:cap} together are equivalent to the inequality $0 < \b < \E [ \th] -  \ubar{\th}$. 

We distinguish the \emph{decision setting}, parameterized by $(F,a,b)$, from the \emph{information technology}, parameterized by $c$. Much of the delegation literature focuses on the uniform--quadratic--constant (UQC) decision setting, where $\th$ is uniformly distributed on the unit interval $[0,1]$; utilities are quadratic; and $b(\th) = \b$ for all $\th$. In this case, Assumptions~\ref{it:f}--\ref{it:cap} hold if and only if $0 < \b < 1/2$. The UQC decision setting will serve as a running example. 


\section{The players' optimization problems} \label{sec:optimization}

Once the principal selects a delegation set, the agent faces a dynamic decision problem. We analyze the agent's problem and then the principal's. 

\subsection{Agent's problem} 

Suppose that the principal has selected a delegation set $D$. We analyze the agent's problem backwards, starting with his choice after each outcome of the experiment. 

\paragraph{After a failure} If the experiment fails, the agent does not learn the state, so he solves
\[
\maz_{y \in D} \quad \E[u_A(y,\theta)].
\] 
The maximum value of this problem, denoted $u_{A,0}(D)$, is called the agent's \emph{uninformed payoff} from $D$. Assume that ties are broken in the principal's favor, and denote the maximizer by $y_{A,0}(D)$, called the agent's \emph{uninformed decision} from $D$. 

\paragraph{After a success} If the experiment succeeds, the agent observes the realized state $\th$, and he solves
\[
\maz_{y \in D} \quad u_A(y,\theta).
\]
Denote the maximum value of this problem by $u_{A} (D, \th)$. Assume that ties are broken in the principal's favor, and denote the maximizer by
$y_A (D, \theta)$, called the agent's \emph{informed decision} from $D$  in state $\th$. The agent's \emph{informed payoff} from $D$ is given by
\[
	u_{A,1} (D) = \E [u_{A}(D,\theta)].
\]

\paragraph{Effort choice} The agent's expected utility gain from observing the state is 
\[
	\D_A(D) = u_{A,1}(D) - u_{A,0}(D).
\]
Learning the state can only help the agent, so $\D_A (D) \geq 0$ for every delegation set $D$. With this notation, the agent's effort choice problem is
\begin{equation} \label{eq:agent_problem}
\maz_{e \in [0,1)} \quad u_{A,0}(D) + e \cdot \D_A(D) -c(e).
\end{equation}
The assumptions on the cost function $c$ ensures that there is a unique maximizer, denoted $\hat{e} (D)$, which is given by the first-order condition
\[
	c'(\hat{e}(D)) = \D_A (D).
\]

%

\subsection{Principal's problem} 

For each delegation set $D$, the principal's expected utility, conditional on each outcome of the experiment, is determined by the agent's subsequent decisions. The principal's uninformed-agent utility and informed-agent utility are defined as
\begin{equation*}
	u_{P,0} (D) = \E [u_P(y_{A,0}(D),\theta)],
	\qquad
	u_{P,1}(D) =\E [u_P(y_A(D, \theta),\theta)].
\end{equation*}
Putting all this together, the principal's delegation problem is
\begin{equation} \label{eq:principal_max}
	\maz_{D} \quad U_P (D)  = (1 - \hat{e}(D)) u_{P,0} (D) + \hat{e} (D) u_{P,1} (D),
\end{equation}
where the maximization is over all compact subsets $D$ of $\R$. The principal's payoff from a delegation set $D$ depends on three terms: her uninformed-agent payoff $u_{P,0} (D)$, her informed-agent payoff $u_{P,1}(D)$, and the agent's expected return $\D_A( D)$ from learning the state, which pins down the agent's effort choice $\hat{e}(D)$. 



The principal's payoff does not change if  she adds to the delegation set superfluous decisions that the agent will never choose. We focus on delegation sets that are \emph{minimal} in the sense that each decision in the delegation set is the agent's uniquely optimal choice in some state.

\begin{lem} [Existence] \label{res:existence} 
	The principal's delegation problem \eqref{eq:principal_max} has a solution. Moreover, there exists an optimal delegation set that is minimal. 
\end{lem}

The principal maximizes over all compact subsets of $\R$. There is no loss in restricting to compact subsets of a fixed, sufficiently large interval. With the Hausdorff metric, this restricted domain is compact. By our tie-breaking assumption, the principal's objective is upper semicontinuous. Thus, a solution exists. 

Hereafter, we restrict attention to delegation sets that are minimal, without explicit reference. In particular, we call a minimal delegation set \textit{the} optimal delegation set if every other \emph{minimal} delegation set is strictly worse.

\section{Optimal delegation} \label{sec:characterization}

\subsection{Informed-agent benchmark}

As a benchmark, consider the classical delegation problem in which the agent knows the state. In our notation, this problem is
\begin{equation} \label{eq:classic_max}
	\maz_{D} \quad u_{P,1}(D),
\end{equation}
where the maximization is over all compact subsets $D$ of $\R$. In this problem, the principal faces a tradeoff between utilizing the agent's private information and restricting the agent's expression of bias. 

For any delegation set $D$, the principal can guarantee the uninformed-agent payoff $u_{P,0} (D)$ by offering the singleton delegation set $\{ y_{A,0} (D) \}$. Therefore, the principal's value from delegating to an informed agent \eqref{eq:classic_max} is an upper bound on the value from delegating to an initially uninformed agent \eqref{eq:principal_max}.

Following \cite{KovacMylovanov2009stochastic} and \cite{Krahmer2016sequential}, we use the envelope theorem to express the principal's objective in \eqref{eq:classic_max} as a weighted average of the \emph{agent's} payoff in each state.

\begin{lem}[Utility representation] \label{res:utility_representation} For any delegation set $D$, we have
\begin{equation} \label{eq:util_representation}
\begin{aligned}
	u_{P,1} (D) 
	&=  u_{A} (D, \ubar{\th}) B (\ubar{\th}) f(\ubar{\th})   -  u_{A} ( D, \bar{\th}) B(\bar{\th}) f(\bar{\th}) \\
	&\qquad + \int_{\ubar{\th}}^{\bar{\th}} u_{A} (D, \th)  [ f(\th) + (B(\th) f (\th))'] \de \th .
\end{aligned}
\end{equation}
\end{lem}

By Assumption \ref{it:f}, the coefficient on $u_A (D, \th)$ in the integral is strictly positive for each state $\th$. By Assumption \ref{it:B}, the coefficient on $u_A (D, \ubar{\th})$ is nonnegative and the coefficient on $u_A ( D, \bar{\th})$ is strictly negative. Call $u_A(D, \th)$ the payoff of type $\th$. As the principal considers enlarging the delegation set, she must balance the gain from increasing the payoff of each type $\th$ in $[\ubar{\th}, \bar{\th})$ against the loss from increasing the payoff of the highest type $\bar{\th}$. The solution gives the agent discretion to choose any decision up to a cap. 




 
The next lemma is equivalent to \citet[Lemma 1, p.~856]{Krahmer2016sequential}.\footnote{\label{ft:weaker}In a quadratic setting with an informed agent, \cite{KovacMylovanov2009stochastic} show that, under their Assumption 1, interval delegation is optimal in the larger class of stochastic mechanisms. Their Assumption 1 is slightly weaker than our assumption \ref{it:f}, which is from \cite{Krahmer2016sequential}. Their Assumption 1 requires the inequality in \ref{it:f} to hold only in states $\th$ for which $y_A(\th)$ lies between the endpoints of the optimal (informed-agent) delegation set. In our main model with information acquisition, we need the stronger assumption  \ref{it:f} because optimal delegation may allow more extreme decisions in order to encourage information acquisition.} 

\begin{lem}[Informed-agent delegation] \label{res:informed_agent_delegation}  The optimal delegation set for an informed agent is the interval $[ y_A ( \ubar{\th}), y_A( \hat{\th})]$, where $\hat{\th}$ is the unique solution of
\begin{equation}
\label{eq:cap_formula}
	\hat{\th} + b(\hat{\th}) = \E [ \th | \th \geq \hat{\th}].
\end{equation}
\end{lem}
Consider the informed agent's choice from the interval $[y_A ( \ubar{\th}), y_A (\hat{\th})]$. If $\th \leq \hat{\th}$, then the agent chooses his favorite decision $y_A( \th)$. If $\th > \hat{\th}$, then the agent chooses the endpoint $y_A ( \hat{\th})$. Condition \eqref{eq:cap_formula} says that the decision $y_A (\hat{\th})$ is the principal's favorite, conditional on the state $\th$ being at least $\hat{\th}$.  Perturbing the cap away from $y_A(\hat{\th})$ perturbs the agent's decision in each state $\th$ with $\th \geq \hat{\th}$. Thus, \eqref{eq:cap_formula} is the first-order condition for the optimal cap. Assumptions \ref{it:f}--\ref{it:cap} ensure that \eqref{eq:cap_formula} has a unique solution. It can be shown that if the bias function $b$ strictly increases pointwise, then the cap $y_A(\hat{\th})$ strictly decreases. Therefore, it is optimal to give an informed agent less discretion if he is more biased.

\subsection{Structure of optimal delegation}

We return to the main delegation problem \eqref{eq:principal_max} in which the agent is initially uninformed. By \cref{res:informed_agent_delegation}, the interval delegation set $[y_A (\ubar{\th}), y_A (\hat{\th})]$ maximizes the principal's informed-agent payoff $u_{P,1}$. As the principal modifies this delegation set, she must trade off the loss from reducing her informed-agent payoff against the potential benefits: increasing her uninformed-agent payoff and inducing greater information acquisition (by decreasing the agent's uninformed payoff  or increasing the agent's informed payoff).

We first characterize the possible ``gaps'' in an optimal delegation set. Formally, a delegation set $D$ \emph{has a gap} $(d_1, d_2)$ if $[d_1, d_2] \cap D = \{ d_1, d_2\}$.\footnote{According to this definition, a proper subset of a gap is not a gap.}


%

\begin{lem}[Gaps] \label{res:gaps}
If an optimal delegation set has a gap $(d_1, d_2)$, then $d_1 < y_{A,0} < d_2$. Moreover, either 
\begin{enumerate}[label= (\roman*), ref = \roman*]
\item \label{it:equal} $\E[u_A ( d_1, \th)] = \E[u_A (d_2, \th)]$; or 
\item  \label{it:unequal} $\E[u_A ( d_1, \th)] > \E[u_A (d_2, \th)]$ and $d_2 > y_A (\bar{\th})$. 
\end{enumerate}
\end{lem}

In words, if an optimal delegation set has a gap, then the gap must contain the agent's ex ante favorite decision $y_{A,0}$. Moreover, the agent must be ex ante indifferent between the endpoints of the gap, unless the right endpoint $d_2$ is strictly above $y_A (\bar{\th})$,  in which case the agent may strictly prefer the left endpoint. 

To prove \cref{res:gaps}, we use the utility representation in \cref{res:utility_representation} to show that the principal strictly prefers to fill (at least partially)  any gap that violates \eqref{it:equal} and \eqref{it:unequal}. The intuition for \cref{res:gaps} is as follows. Consider a delegation set $D$ satisfying $D \subset [y_A (\ubar{\th}), y_A(\bar{\th})]$. Gaps in $D$ clearly reduce the agent's informed payoff. By Assumption~\ref{it:f}, gaps in $D$ also reduce the principal's informed-agent payoff. Therefore, any benefit from a gap must come from its effect on the agent's uninformed decision. If a delegation set $D$ is optimal, then it must have the minimal gap necessary to induce the uninformed decision $y_{A,0} (D)$. Such a gap contains exactly those decisions that the agent strictly prefers to $y_{A,0}(D)$. Hence, \eqref{it:equal} follows. This argument does not apply to gaps $(d_1, d_2)$ with $d_2 > y_A ( \bar{\th})$. Filling in decisions near $d_2$ can reduce the principal's informed-agent payoff. In some states, the agent switches his decision from $d_1$ to the higher, newly available decisions. The agent's decision decreases only in those extremely high states in which $d_2$ was chosen. We will give an example of an optimal delegation set with a gap that satisfies \eqref{it:unequal}.

Using \cref{res:gaps}, we now characterize the three forms that optimal delegation can take. Recall the definition of the cutoff type $\hat{\th}$ from the optimal informed-agent delegation set (\cref{res:informed_agent_delegation}).


\begin{thm}[Optimal delegation---three forms] \label{res:characterization}
	If a delegation set $D^\ast$ is optimal, then one of the following holds.
	\begin{enumerate}[label = \Roman*., ref = \Roman*]
		\item \label{it:hollow} Hollow: $D^\ast = [y_A  (\ubar{\th}) \wedge y_0, y_0] \cup [y_1, y_2]$ for some $y_0$, $y_1$, and $y_2$ satisfying $y_0 < y_{A,0} < y_1 \leq y_2$, where $\E [u_A (y_0, \th)] = \E [u_A ( y_1, \th)]$ and $y_2 > y_A (\hat{\th})$.
		\item \label{it:interval} Interval: $D^\ast  = [y_A  ( \ubar{\th}), y_0]$ for some $y_0$ satisfying $y_{P,0} < y_0 < y_{A,0}$. 
		\item \label{it:dominated} High-point: $D^\ast = [ y_A ( \ubar{\th}) \wedge y_0, y_0] \cup \{ \bar{y}\}$ for some $y_0$ and $\bar{y}$ satisfying $y_0 < y_{A,0} < y_A ( \bar{\th}) < \bar{y}$  and $\E[u_A ( y_0, \th)] > \E [ u_A ( \bar{y}, \th)]$. 
	\end{enumerate}
\end{thm}


\def\r{0.14}
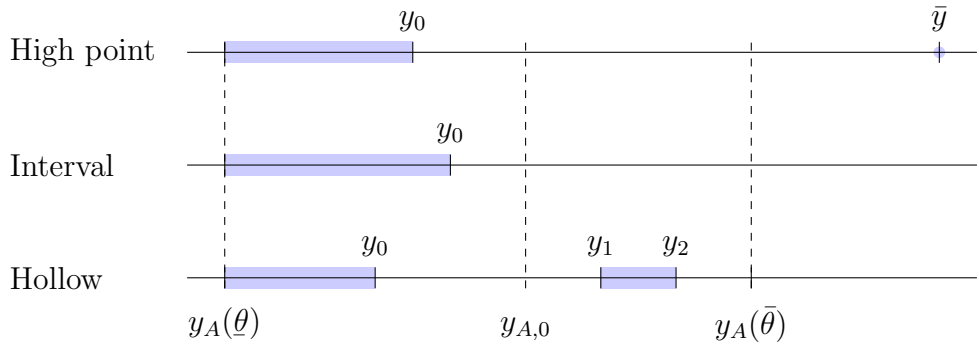
\begin{figure}
\begin{center}
	\begin{tikzpicture}

		\fill [opacity = 0.2, blue] (1, 1.5 -\r) -- (1, 1.5 + \r) -- (4, 1.5 + \r) -- (4, 1.5 -\r) -- cycle;
		\fill [opacity = 0.2, blue] (1, - \r) -- (1, \r) -- (3, \r) -- (3, -\r) -- cycle; 
		\fill [opacity = 0.2, blue] (6,  - \r) -- (6, \r) -- (7, \r) -- (7, -\r) -- cycle;
		\fill [opacity = 0.2, blue] (1, 3 - \r) -- (1, 3+\r) -- (3.5, 3 + \r) -- (3.5, 3 -\r) -- cycle;
		\filldraw [opacity = 0.2, blue] (10.5, 3 ) circle (2pt);
		
		\draw (-2,0) node[anchor = west]  {Hollow};
		\draw (-2,1.5) node [anchor  = west] {Interval};
		\draw (-2, 3) node [anchor = west]  {High point~};
		
		\draw (1,\r) -- (1,-\r);
		\draw (1, -0.25) node [below] {$y_A ( \ubar{\th})$};
		\draw [dashed] (1, -\r) -- (1, 3 + \r);
		\draw (5, -0.25) node [below]  {$y_{A,0} \vphantom{y_A ( \ubar{\th})}$};
		\draw [dashed] (5, -\r) -- (5, 3 + \r ); 
		\draw (8,\r) -- (8,-\r);
		\draw (8, -0.25) node [below] {$y_A (\bar{\th})$};
		\draw [dashed] (8, -\r) -- (8, 3 + \r ); 
		
		\draw (11,0) -- (0.5,0);
		\draw (11,1.5) -- (0.5,1.5); 
		\draw (11,3) -- (0.5,3);
		\draw (4,1.5-\r) --  (4,1.5 + \r) node [above] {$y_0$};

		\draw (1,1.5-\r) --  (1,1.5 + \r);
		\draw (1,3-\r) --  (1,3 + \r);
		\draw (3,-\r) --  (3,\r) node [above] {$y_0$};
		\draw (6,-\r) --  (6,\r) node [above] {$y_1$};
		\draw (7,-\r) --  (7,\r) node [above] {$y_2$};
		\draw (3.5,3-\r) --  (3.5,3 + \r) node [above] {$y_0$};
		\draw (10.5,3-\r) --  (10.5,3 + \r) node [above] {$\bar{y}$};
		
	\end{tikzpicture}
\end{center}
	\caption{Three delegation forms}
	\label{fig:delegation_forms}
\end{figure}

\cref{fig:delegation_forms} shows an example of each of the three delegation forms. In every case, the agent's ex ante favorite decision $y_{A,0}$ is excluded. This is because the only first-order effect of hollowing out a small gap around $y_{A,0}$ is to increase the principal's uninformed-agent payoff. Recall that in our setting, the  optimal informed-agent delegation set is an interval.\footnote{The optimal informed-agent delegation set can be quite irregular if the state distribution violates our assumption \ref{it:f}; see \citet[Proposition 2, p.~273]{alonso2008optimal}.  If the density decreases too quickly, then the optimal delegation set may have gaps in order to deter deviations by lower types. In our model, there is a positive probability that the agent's posterior mean equals the prior mean. Therefore, the density effectively decreases quickly to the right of the prior mean. The new effect in our setting is the endogeneity of the agent's type distribution.} Starting from the optimal informed-agent delegation set $[y_A(\ubar{\th}), y_A(\hat{\th})]$, a \emph{hollow} delegation set is formed as follows. First, the cap is increased from $y_A(\hat{\th})$ to $y_2$. Raising the cap now has the additional benefit of increasing information acquisition. Second, the principal hollows out a gap $(y_0, y_1)$ around the agent's ex ante favorite decision. If the experiment fails, the agent is indifferent between $y_0$ and $y_1$, but he chooses the lower decision $y_0$ by the tie-breaking assumption. The gap reduces both the agent's informed and uninformed payoffs, but it reduces the agent's uninformed payoff by more---the agent's uninformed decision changes in every state, but his informed decision changes only in some states. Therefore, enlarging the gap increases information acquisition, but it also decreases the principal's informed-agent payoff.\footnote{Enlarging the gap also affects the principal's uninformed-agent payoff. The sign of this effect depends on the relative values of $y_0$ and $y_{P,0}$.}

 If the agent's bias is large enough, then allowing decisions above the agent's ex ante favorite decision $y_{A,0}$ entails a direct loss that outweighs the information acquisition benefit. In this case, an \emph{interval} delegation set is optimal. Each interval delegation set includes the principal's ex ante favorite decision but excludes the agent's ex ante favorite decision.
 
 
A \emph{high-point} delegation set consists of an interval together with an isolated point $\bar{y}$ that is strictly higher than even the agent's favorite decision in the highest state.  If the experiment fails, the agent strictly prefers $y_0$ to the high point $\bar{y}$. This strict preference distinguishes high-point delegation from the special case of hollow delegation with $y_1 = y_2$. The gap $(y_0, \bar{y})$ is consistent with \cref{res:gaps} because $\bar{y} > y_A (\bar{\th})$.  The optimal value of $\bar{y}$ balances two competing forces. Increasing the high point discourages information acquisition by decreasing the agent's informed payoff, while leaving the agent's uninformed payoff unchanged. On the other hand, perturbing the high point below its optimal value reduces the principal's informed-agent payoff by inducing the agent to change his decision from $y_0$ to the new, lower high point in certain states.\footnote{The principal's informed-agent payoff is a qasiconvex function of the high-point (\cref{res:effect_parameters}).  At the optimal high-point, this function must be locally strictly increasing; otherwise the principal would profit from reducing the high point.} 

We claim that high-point delegation is never optimal in the related models of \cite{szalay2005economics} and \cite{semenov2018delegation}. Consider an arbitrary high-point delegation set $D$. If the agent is unbiased, as in \cite{szalay2005economics},\footnote{\cite{szalay2005economics} restricts the principal to delegation sets within $[y_A (\ubar{\th}), y_A (\bar{\th})]$, so high-point delegation is not feasible. Our argument above shows that if this restriction is dropped, high-point delegation cannot be optimal.} then the principal strictly prefers to perturb the high-point downward, thus increasing the common informed-agent payoff $u_{P,1} = u_{A,1}$ and inducing greater information acquisition. If the agent is biased but exogenously informed of the state with a fixed probability, as in \cite{semenov2018delegation},  then the principal's payoff is a fixed convex combination of her uninformed-agent and informed-agent payoffs. Perturbing the high point affects only the principal's informed-agent payoff $u_{P,1}$. As a function of the high-point, this payoff $u_{P,1}$ is strictly quasiconvex (see \cref{res:effect_parameters}) and hence cannot have an interior maximizer. 




\subsection{Optimal delegation form}

The characterization in \cref{res:characterization} reduces the class of candidate optimal delegation sets to a small parametric family. In any example, we can numerically optimize over the parameters to find an optimal delegation set. We now give general conditions under which the different delegation forms are optimal. 

Recall that the agent's optimal effort level $\hat{e} (D)$ depends only on the agent's gain $\D_A(D)$ from learning the state. Let $\hat{e} (x)$ denote the agent's optimal effort level if $\D_A(D) = x$. Formally, $\hat{e}(x)$ is defined by $c'(\hat{e}(x)) = x$ for any $x > 0$. Below, the context should make clear whether the argument of $\hat{e}$ is a delegation set or a number. 


\begin{thm}[Optimal delegation form]  \label{res:delegation_regime}
	Fix a decision setting $(F, a, b)$. 
	\begin{enumerate}
		\item \label{it:low_bias} If  $\E [ \th + b(\th)] \leq \hat{\th} + b(\hat{\th})$,  then for every cost function, every optimal delegation set is hollow. 
		\item \label{it:high_bias} If  $\E [ \th + b(\th)] > \hat{\th} + b(\hat{\th})$, then there exists a positive threshold $K = K (F,a, b)$ such that if $-\hat{e}''(x) / \hat{e}'(x) \geq K$ for all positive $x$, then every optimal delegation set is an interval.
		\item \label{it:very_high_bias} If $u_{P,1} ( \{ y_{P,0}\})  \geq  u_{P,1} ( [ y_A (\ubar{\th}), y_{A,0}])$, then for every cost function, every optimal delegation set is either an interval or high-point. 
	\end{enumerate}
\end{thm}

The inequality $\E[ \th + b(\th)] \leq \hat{\th} + b(\hat{\th})$ is a low-bias condition.\footnote{Formally, if this inequality holds for some bias function $b$, then it can be shown to hold for any pointwise smaller bias function.} It holds if and only if the agent's ex ante favorite decision $y_{A,0}$ lies inside the optimal informed-agent delegation set $[y_A(\ubar{\th}), y_A (\hat{\th})]$ from \cref{res:informed_agent_delegation}. In this case, for any interval delegation set, the principal strictly prefers to add some decision above $y_{A,0}$ that does not change the agent's uninformed decision. This modification increases the principal's informed-agent payoff and encourages information acquisition. Thus, hollow delegation is optimal (part \ref{it:low_bias}). If instead $y_{A,0}$ lies outside $[y_A(\ubar{\th}), y_A (\hat{\th})]$, then allowing such decisions above $y_{A,0}$ necessarily reduces the principal's informed-agent payoff. If
$\hat{e}$ is sufficiently concave, then this loss cannot be outweighed by the information acquisition benefit. Hence, interval delegation is optimal (part \ref{it:high_bias}). Finally, for very high bias, hollow delegation cannot be optimal, no matter the cost function (part \ref{it:very_high_bias}).  In this case, the principal prefers retaining decision control (by selecting the singleton delegation set $\{ y_{P,0}\}$) to any hollow delegation set. 

We illustrate  \cref{res:delegation_regime} in the UQC setting with effort function $\hat{e} (x) = 1 - e^{-x/\k}$, where $\k > 0$. This effort cost function is convenient because the Arrow--Pratt coefficient $-\hat{e}''(x) /\hat{e}'(x) $ equals $1/\k$ for all $x$. The function $\hat{e}$ is induced by the cost function
\begin{equation} \label{eq:Szalay_cost}
	c(e) = \k [ (1 - e) \log (1 - e) + e],
\end{equation}
which is the leading example in \cite{szalay2005economics}. This UQC-exponential setting is parametrized by the bias parameter $\b$ and the cost parameter $\k$.

\begin{figure}
\begin{center}
\includegraphics[scale=0.7]{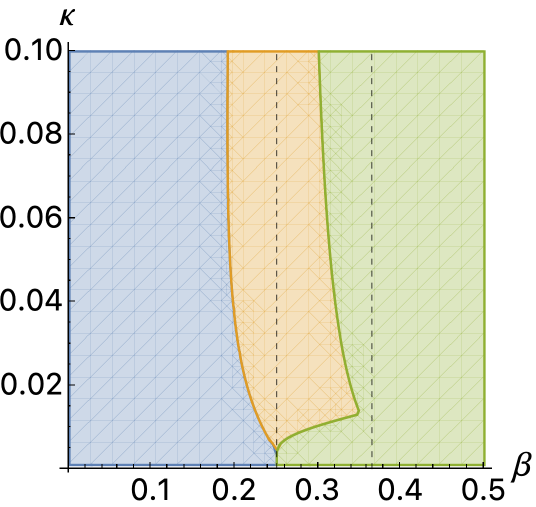}
\qquad
\includegraphics[scale=0.7]{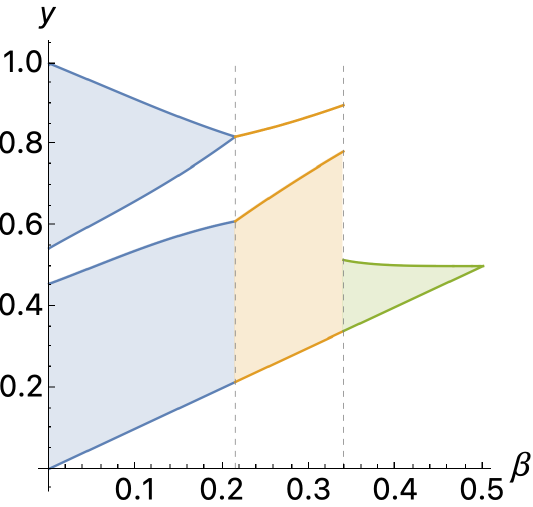}
\end{center}
\caption{Optimal delegation forms}
\label{fig:optimal_form}
\end{figure}

 The left panel of  \cref{fig:optimal_form} indicates the optimal form of delegation as a function of the parameter vector $(\b, \k)$. In this parametric setting, high-point delegation is not optimal. The three regions (blue, orange, green) respectively indicate where the following delegation forms are optimal: hollow with non-degenerate upper interval $[y_1, y_2]$, hollow with singleton upper interval, and interval. In particular, the interval delegation region is not convex. For some values of $\b$, hollow delegation is optimal only for intermediate values of $\k$. The upper interval of the hollow delegation set encourages information acquisition. For low $\k$, information acquisition is high without the upper interval. For high $\k$, information acquisition is low even with the upper interval. 
 
The conditions in \cref{res:delegation_regime} have simple expressions in terms of $(\b, \k)$. By part \ref{it:low_bias}, hollow delegation is optimal if $\b \leq 1/4$, i.e., left of the first dotted line. By part \ref{it:high_bias}, for bias $\b > 1/4$, hollow delegation is optimal if $\k \leq 1 / K(\b)$, where $K$ is a positive function. By part \ref{it:very_high_bias}, interval (or high-point) delegation is optimal if $\b \geq (\sqrt{33} + 3)/8 \approx 0.36$, i.e., right of the second dotted line. 

The right panel of \cref{fig:optimal_form} plots the optimal delegation set, as a function of the bias $\b$, for fixed $\k  = 0.02$. This graph visualizees the optimal delegation sets for parameter values along the horizontal line $\k=0.02$ in the left panel. As the bias increases from $\b = 0$, the top interval of the optimal hollow delegation set shrinks until it becomes a point. Thereafter, the upper interval remains degenerate until the optimal delegation form switches from hollow to interval. At this transition, information acquisition jumps down, and the principal's uninformed-agent payoff jumps up. 

It is difficult to give general sufficient conditions for high-point delegation to be optimal. In the UQC decision setting, for each bias $\b$ in $(1/4, 1/2)$ we construct a cost function for which high-point delegation is optimal; see \cref{sec:ex_high_point}. In our construction, the effort function $\hat{e}$ is a differentiable, strictly increasing approximation of a step function: $\hat{e} ( x) \approx  \e [ x \geq x_0]$, where $x_0$ is  slightly larger than $\D_A( [ y_A(\ubar{\th}), 1/2])$, and $\e$ is a sufficiently small positive parameter. For any delegation set $D$, the agent's experiment fails with high probability. Therefore, a necessary condition for a delegation set to be optimal is that it must induce an uninformed decision near $y_{P,0} = 1/2$. Among delegation sets satisfying this necessary condition, intervals do not induce sufficient information acquisition, and hollow sets result in high decisions that bring down the principal's informed-agent payoff. The principal strictly prefers a high-point delegation set. The high point encourages information acquisition (by pushing the agent's return from learning the state above $x_0$), while preserving the principal's informed-agent payoff (by ensuring that decisions above $y_{A,0}$ are selected with low probability).

\section{Principal's preferences over the agent} \label{sec:bias}

Suppose that the principal initially selects an agent from a pool of candidates with different information acquisition technologies and different biases. Whom should the principal select?  Of course, the principal's first choice would be an unbiased agent who can perfectly learn the state for free, but such an agent may not be available. It is intuitive that the principal would prefer cheaper information acquisition and lower bias. Cheaper information is indeed better for the principal, but we show that reducing the agent's bias can hurt the principal. 


Consider an optimal delegation set $D^\ast$ for a fixed cost function $c$. If the marginal cost of information acquisition, $c'$, decreases pointwise, then the agent's induced effort choice function $\hat{e}$ increases pointwise. In particular, $\hat{e} ( D^\ast)$ increases, so the principal's payoff from $D^\ast$ increases,\footnote{By Assumption~\ref{it:cap}, a singleton delegation set is suboptimal, so we must have  $u_{P,1}(D^\ast) > u_{P,0} (D^\ast)$.} and hence the principal's value from delegation must increase. 

We now turn to the comparative statics in the agent's bias. To build intuition, consider the UQC setting. Fix an arbitrary delegation set $D$. As the agent's bias $\b$ decreases, the principal's informed-agent payoff $u_{P,1}(D)$ and uninformed-agent payoff $u_{P,0}(D)$ both increase. The effect on the agent's return $\D_A(D)$ from learning the state depends on the form of $D$. If $D$ is an interval delegation set, then $\D_A(D)$ also increases, and therefore the principal strictly benefits. For hollow delegation sets, however, the agent's return from learning the state can decrease. The resulting loss from reduced information acquisition can outweigh the benefit of less biased decisions after each outcome of the experiment.

\begin{thm}[Benefiting from bias]  \label{res:CS}
	Assume that the players have quadratic-loss utilities and the state distribution is symmetric about its mean. Let $e_0$ denote the level of effort chosen by an unbiased agent faced with an unrestricted delegation set. If
	\begin{equation} \label{eq:condition_Szalay}
		\frac{ c'(e_0)}{c''(e_0)} > 1 - e_0, 
	\end{equation}
	then there exists $\bar{\b} > 0$ such that for all $\b$ in $(0, \bar{\b})$, the principal strictly prefers an agent with constant bias $\b$ to an unbiased agent. 
\end{thm}

In the application of a politician delegating to a policy advisor who conducts costly research, \cref{res:CS} indicates that the politician may be better off selecting an advisor whose political preferences differ from her own. The intuition for \cref{res:CS} is as follows. If the agent is unbiased, then the principal and agent agree on the ex ante optimal decision. Removing decisions near this ex ante optimum encourages information acquisition by punishing the agent if his experiment fails. But this punishment for the agent also punishes the principal. If the agent is biased, then the principal and agent prefer different decisions ex ante. The principal can therefore select a delegation set that excludes the agent's ex ante favorite decision but includes her own. This intuition relies on the property of the information acquisition technology that, with positive probability, the agent learns little about the state. On the other hand, we believe that the potential benefit from bias is robust to relaxing the assumptions that the agent perfectly learns the state with positive probability.




By \cref{res:delegation_regime}.\ref{it:low_bias}, optimal delegation is hollow if the agent's bias is sufficiently small (and nonzero). In the limit as the bias tends to zero, the hollow delegation gap may or may not vanish, depending on the cost of information acquisition. By \citet[Proposition 3, p.~1181]{szalay2005economics},  if condition \eqref{eq:condition_Szalay} holds, then with an unbiased agent, any optimal delegation set $D^\ast$ must have a gap around the ex ante favorite decision $y_{P,0} = y_{A,0}$. Hence, the agent's uninformed decision is strictly below $y_{P,0}$. We show that for any sufficiently small bias $\b$, the principal's payoff from offering the shifted delegation set $\b + D^\ast$ to an agent with bias $\b$ is strictly higher than the principal's payoff from offering the delegation set $D^\ast$ to an unbiased agent. The first-order benefit from inducing a less biased uninformed decision outweighs the second-order loss from inducing more biased decisions when the agent learns the state. As the bias $\b$ varies, shifting the delegation set in this way keeps constant the agent's return from learning the state, and hence the induced effort level.

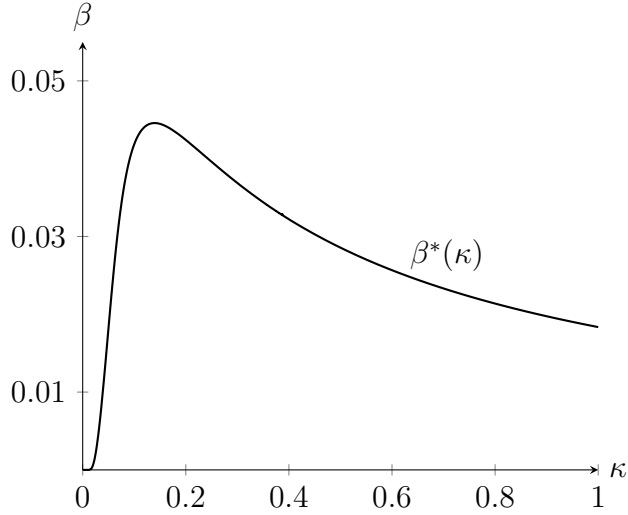
\begin{figure}
	\begin{center}
		\begin{tikzpicture}
			\begin{axis}[
				clip = false,
				scaled y ticks = false,
				axis y line =middle, 
				axis x line=bottom,
				clip = false,
				xmin = 0,
				xmax = 1,
				ymin = 0,
				ymax = 0.055,
				ytick = {0.01, 0.03, 0.05},
				yticklabels = {0.01,0.03, 0.05},
				xlabel = {$\k$},
				ylabel = {$\b$},
				x label style={at={(current axis.right of origin)},anchor=west},
				y label style = {at={(current axis.above origin)},anchor=south},]
				]
				\addplot [thick, line join = bevel] table[col sep=comma]{optimal_bias.dat} node [pos = 0.7, above, xshift = 2pt, yshift = 2pt] {$\b^\ast (\k)$};
		\end{axis}
			
		\end{tikzpicture}
	\end{center}
	\caption{Principal's favorite bias}
	\label{fig:benefitig}
\end{figure}

To illustrate \cref{res:CS}, we return to the UQC setting with the cost function from \eqref{eq:Szalay_cost}. We have $c'(e_0)/ c''(e_0)  = - (1 - e_0) \log (1 - e_0) $ The effort level $e_0$ depends on $\k$. It can be checked that \eqref{eq:condition_Szalay} holds if and only if $\k \leq 1/12$.  \cref{fig:benefitig} plots the principal's favorite level of agent bias, $\b^\ast (\k)$, as a function of the cost parameter $\k$. The principal's favorite bias achieves a maximum value of approximately $0.045$. In the UQC setting with an informed agent, it is optimal for the principal to delegate if and only if $\b \leq 0.5$. So among the range of biases for which delegation is worthwhile, the bottom $9\%$ are more aligned than is optimal. The principal's gain from an optimally biased agent relative to an unbiased agent peaks between $\k = 0.07$ and $\k = 0.08$. With $\k = 0.08$, the principal's gain from facing an agent with the optimal bias $\b^\ast (0.08) \approx 0.036$ (rather than an unbiased agent) is equivalent to her loss, in the informed-agent problem, from facing an agent with bias $\b = 0.027$ (rather than an unbiased agent).

\section{Conclusion} \label{sec:conclusion}




In the classical delegation literature, the agent's bias is all that prevents the principal from achieving her first-best decision rule. We consider an additional friction---costly information acquisition for the agent. We characterize the principal's optimal delegation set, and we show that the principal may prefer delegating to an agent with a small bias than to an unbiased agent. Preference misalignment allows the principal to punish the agent at a lower cost to herself. These punishments occur on-path because the agent cannot learn the state with certainty. 

We work with a flexible decision setting, but the information acquisition technology takes a simple parametric form. This structure allows us to characterize the optimal delegation set.  Allowing flexible information acquisition presents a few challenges. In our model, the agent's optimal effort choice depends on a simple statistic of the delegation set, and the resulting distribution of the agent's posterior beliefs lies in a one-parameter family. With flexible information acquisition, the agent's optimal choice of information structure depends on the delegation set in a complicated way, and there are no exogenous restrictions on the resulting distribution of the agent's posterior beliefs. Incorporating flexible information acquisition into models of delegation is a promising direction for future work. 



\appendix

\newpage

\section{Proofs} \label{sec:proofs}

In the proofs, we use the following terminology. Given a state $\th$ and a decision $y$  with $y\neq y_A (\th)$, the \emph{$\th$-conjugate} of $y$ is the unique decision $y'$ distinct from $y$ that satisfies $u_A ( y', \th) = u_A ( y, \th)$. Decision $y_A(\th)$ is defined to be the $\th$-conjugate of itself. Similarly, given a decision $y$ with $y \neq y_{A,0}$, the \emph{ex ante conjugate} of $y$ is the unique decision $y'$ distinct from $y$ that satisfies $\E [u_A (y, \th)] = \E [u_A (y', \th)]$. Decision $y_{A,0}$ is defined to be the ex ante conjugate of itself. 

\subsection{Proof of \texorpdfstring{\cref{res:existence}}{Proposition \ref{res:existence}}} \label{sec:proof_existence}

\paragraph{Existence} Select a sufficiently large compact set $Y \subseteq \R$ so that restricting to compact subsets of $Y$ does not change the supremum. Denote by $\KK_Y$ the space of nonempty compact subsets of $Y$, endowed with the Hausdorff metric. This space $\KK_Y$ is compact \cite[Theorem 3.85]{AliprantisBorder2006}. To prove existence, it suffices to check that the map
\[
D \mapsto (1 - \hat{e}(D)) u_{P,0} (D) + \hat{e}(D) u_{P,1} (D)
\]
is upper semicontinuous. We prove that $u_{P,0}$ and $u_{P,1}$ are upper semicontinuous and $\hat{e}$ is continuous. 

Define correspondences $Y_0^\ast \colon \KK_Y \twoheadrightarrow Y$ and $Y_1^\ast \colon \KK_Y \times [\ubar{\th}, \bar{\th}] \twoheadrightarrow Y$ by
\[
Y_0^\ast (D) = \argmax_{ y \in D}\, \E [ u_A ( y, \th)], \qquad Y_1^\ast (D, \th) = \argmax_{ y \in D}\,  u_A( y, \th).
\]
Endow $\KK_Y \times [\ubar{\th}, \bar{\th}]$ with the product topology. By Berge's theorem \cite[Theorem 17.31]{AliprantisBorder2006}, these correspondences are upper hemicontinuous.\footnote{The function $y \mapsto \E [u_A ( y, \th)]$ is continuous by dominated convergence. For $Y_0^\ast$, we apply Berge's theorem to the correspondence $\varphi_0$ from $\KK_Y$ into $Y$ defined by $\varphi_0 (D) = D$. The associated identity function on $\KK_Y$ is clearly continuous, so $\varphi_0$ is continuous by Theorem 17.15 in \cite{AliprantisBorder2006}. For $Y_1^\ast$, we apply Berge's theorem to the correspondence $\varphi_1$ from $\KK_Y \times [\ubar{\th}, \bar{\th}]$ into $Y$ defined by $\varphi_1 (D, \th) = D$. This correspondence is continuous because it is the composition of $\varphi_0$ with the projection map $(D, \th) \mapsto D$.}
By our tie-breaking assumption, the utility functions $u_{P,0}$ and $u_{P,1}$ can be expressed as
\[
u_{P,0} (D) = \max_{y \in Y_0^\ast (D)} \E[ u_P (y, \th)]
\quad
\text{and}
\quad	
u_{P,1} (D) = \E \Brac{ \max_{y \in Y_1^\ast (D, \th)} u_P(y, \th)}.
\]
By a variant of Berge's theorem \cite[Lemma 17.30]{AliprantisBorder2006}, it follows that $u_{P,0}$ and $u_{P,1}$ are upper semicontinuous.%
\footnote{For $u_{P,0}$, conclude by dominated convergence that the function $y \mapsto \E [ u_P (y, \th)]$ is continuous, hence upper semicontinuous. For $u_{P,1}$, Berge's theorem implies that the integrand is upper semicontinuous in $(D,\th)$, so $u_{P,1}$ is upper semicontinuous by dominated convergence.}

Now we check that $\hat{e}$ is continuous. Recall that $\D_A (D) = u_{A,1} (D) - u_{A,0}(D)$. It follows, as above, from Berge's theorem \cite[Theorem 17.31]{AliprantisBorder2006} that $\D_A$ is a continuous function on $\KK_Y$. The agent's effort choice first-order condition gives $\hat{e}(D) = (c')^{-1} ( \D_A (D))$. The inverse function $(c')^{-1}$ is well-defined and continuous because $c'$ is strictly increasing, with $c'(0) = 0$ and $\lim_{e \uparrow 1} c'(e) = \infty$. Therefore, $\hat{e}$ is the composition of continuous functions, and hence is continuous. 

\paragraph{Minimality} Formally, a delegation set $D$ is \emph{minimal} if for each decision $d$ in $D$ there exists some state $\th$ in $[\ubar{\th}, \bar{\th}]$ such that $\{d\} = \argmax_{y \in D } u_A ( y, \th)$. 

Now we turn to the proof. Given an arbitrary compact delegation set $D$, let 
\[
	\ubar{y} = \max \Brac{ \argmax_{y \in D}\, u_A ( y, \ubar{\th})},
	\qquad
	\bar{y} = \min \Brac{ \argmax_{y \in D}\, u_A (y, \bar{\th})}.
\]
Let $D' = [ \ubar{y}, \bar{y}] \cap D$. By construction, $D'$ is minimal.\footnote{Fix $d$ in $D'$. If $d$ is in $[y_A(\ubar{\th}), y_A ( \bar{\th})]$, set $\th = y_A^{-1} ( d)$. If $d  < y_A (\ubar{\th})$, set $\th = \ubar{\th}$. If $d > y_A (\bar{\th})$, set $\th =  \bar{\th}$. In each case, $\{d\} = \argmax_{ y \in Y} u_A ( y, \th)$.} The agent makes the same choice from $D'$ as from $D$, unless he learns that the state is in $\{ \ubar{\th}, \bar{\th}\}$, which occurs with probability $0$. Therefore, $U_P(D') = U_P(D)$.

\subsection{Proof of \texorpdfstring{\cref{res:utility_representation}}{Proposition \ref{res:utility_representation}}}

Fix a delegation set $D$. By the envelope theorem \citep{MilgromSegal2002},\footnote{The conditions of the envelope theorem are satisfied because the partial derivative $D_\th u_A (y, \th)$ equals $y (1 + b'(\th))$, which is bounded by $(\max |D|) (1 + \max_{\xi \in [\ubar{\th}, \bar{\th}]} b'(\xi))$, which is finite because $b$ is continuously differentiable.} we have
\begin{equation} \label{eq:envelope_integration}
	u_A (D, \th) - u_A ( D, \ubar{\th}) = \int_{\ubar{\th}}^{\th} (1 + b'(\xi)) y_A ( D, \xi) \de \xi, 
\end{equation}
for each state $\th$. Since
\[
	u_P (y, \th) -  u_A (y, \th) = - b(\th) y = - (1 + b'(\th)) B(\th) y,
\]
we also have
\[
	u_{P,1} (D) - u_{A,1} ( D)  = - \int_{\ubar{\th}}^{\bar{\th}} (1 + b'(\th))  y_A (D, \th) B(\th) f(\th) \de \th. 
\]
Now integrate by parts, using $u_A(D, \th)$ as an antiderivative, from \eqref{eq:envelope_integration}, to get
\[
u_{P,1} (D) - u_{A,1} (D)  =  - u_A ( D, \th) B(\th) f(\th) \big\vert_{\ubar{\th}}^{\bar{\th}}   + \int_{\ubar{\th}}^{\bar{\th}} u_A(D, \th)  (B (\th) f(\th))' \de \th. 
\]
Bringing $u_{A,1} (D)$ to the right side and expressing it as integral gives \eqref{eq:util_representation}. 

\subsection{Proof of \texorpdfstring{\cref{res:informed_agent_delegation}}{Proposition \ref{res:informed_agent_delegation}}} \label{sec:proof_informed_agent_delegation}

Let $D$ be a maximizer of $u_{P,1}$ that is minimal, which exists by the proof of \cref{res:existence} (\cref{sec:proof_existence}). By \ref{it:cap}, we know 
$y_{P,0} > y_{A} (\ubar{\th})$, so  $\max D > y_A ( \ubar{\th})$. We also claim that $\max D \leq y_A ( \bar{\th})$. Otherwise, letting $\bar{d}$ be the $\bar{\th}$-conjugate of $\max D$,\footnote{See the beginning of \cref{sec:proofs} for the definition.} the principal would strictly prefer $[y_A(\ubar{\th}) \wedge \bar{d}, \bar{d}]$ by the utility representation (\cref{res:utility_representation}).
 
We have shown that $y_{A} (\ubar{\th}) < \max D \leq y_A ( \bar{\th})$, so  $\max D = y_A ( \hat{\th})$ for some $\hat{\th}$ in $(\ubar{\th}, \bar{\th}]$. Hence, $ D = [y_A (\ubar{\th}), y_A (\hat{\th})]$, for otherwise the principal would strictly prefer $[y_A (\ubar{\th}), y_A (\hat{\th})]$ by the utility representation (\cref{res:utility_representation}). It remains to maximize over the value of $\hat{\th}$. We have
\[
	u_{P,1} ( [y_A (\ubar{\th}), y_A (\hat{\th})]) = \int_{\ubar{\th}}^{\hat{\th}} u_P (y_A (\th), \th) f(\th) \de \th + \int_{\hat{\th}}^{\bar{\th}} u_P (y_A ( \hat{\th}), \th) f(\th ) \de \th. 
\]
Differentiate with respect to $\hat{\th}$ and note that $a'(y_A (\hat{\th})) = - \hat{\th} - b(\hat{\th})$, to get 
\begin{equation} \label{eq:deriv}
	y_A'(\hat{\th}) \int_{\hat{\th}}^{\bar{\th}} [ \th  - \hat{\th} - b(\hat{\th}) ] f(\th ) \de \th.
\end{equation}

Since $y_A'$ is strictly positive, we focus on the sign of the integral, as a function of $\hat{\th}$. It is positive at $\hat{\th} = \ubar{\th}$ since $\E[\th] - \ubar{\th} - b(\ubar{\th}) > 0$ (by \ref{it:cap}), and it becomes negative near $\bar{\th}$ since $-b(\bar{\th}) < 0$ (by \ref{it:B}). We now check that the integral in \eqref{eq:deriv} is strictly single-crossing from above. Differentiating gives
\[
	b(\hat{\th}) f(\hat{\th}) - (1 + b' (\hat{\th}))(1 - F(\hat{\th}))
	= - (1 + b'(\hat{\th}))( 1 - F(\hat{\th}) - B(\hat{\th}) f(\hat{\th})).
\]
This expression is in turn single-crossing from above since $1 + b'(\hat{\th}) > 0$ and $1 - F(\hat{\th}) - B(\hat{\th}) f(\hat{\th})$ is strictly decreasing (by \ref{it:f}). It follows that the integral in \eqref{eq:deriv} is also strictly single-crossing from above.

\subsection{Proof of \texorpdfstring{\cref{res:gaps}}{Proposition \ref{res:gaps}}} \label{sec:proof_gaps}

Let $D^\ast$ be an optimal delegation set that is minimal. We must have $\D_A ( D^\ast) > 0$ and $u_{P,1} (D^\ast) > u_{P,0} (D^\ast)$; otherwise, the principal would strictly prefer the delegation set $[y_A ( \ubar{\th}), y_{P,0}]$ since $ y_A ( \ubar{\th}) < y_{P,0} < y_A ( \hat{\th})$ by \ref{it:cap} and \cref{res:informed_agent_delegation}.

Suppose, for a contradiction,  that $D^\ast$ has a gap $(d_1, d_2)$ that violates \eqref{it:equal} and \eqref{it:unequal}. There are four cases. 
\begin{enumerate}
	\item $y_{A,0} \not\in (d_1, d_2)$ and $d_2 \leq y_A (\bar{\th})$. Take $D' = D^\ast \cup (d_1, d_2)$. 
	
	\item $y_{A,0} \not\in (d_1, d_2)$ and $d_2 >  y_A (\bar{\th})$. By minimality, $\{ d_2\} = \argmax_{y \in D} u_A ( y, \bar{\th})$. In particular, $u_A ( d_2, \bar{\th}) > u_A ( d_1, \bar{\th})$. Let $d_1'$ be the $\bar{\th}$-conjugate of $d_2$. Take $D' = D^\ast \cup (d_1, d_1']$. 
	
	\item $y_{A,0} \in (d_1, d_2)$ and  $\E [ u_A (d_1, \th)] < \E[ u_A (d_2, \th)]$. Let $d_1'$ be the ex ante conjugate of $d_2$. Take $D' = D^\ast \cup (d_1, d_1']$. 
	
	\item $y_{A,0} \in (d_1, d_2)$ and  $\E [ u_A (d_1, \th)] > \E[ u_A (d_2, \th)]$ and $d_2 \leq y_A(\bar{\th})$. Let $d_2'$ be the ex ante conjugate of $d_1$. Take $D' = D^\ast \cup [d_2', d_2)$. 
\end{enumerate}

In every case, we have $u_{P,0} (D') \geq u_{P,0}(D^\ast)$; $\D_A( D') > \D_A (D^\ast)$; and $u_{P,1} (D') > u_{P,1}(D^\ast)$ by the utility representation in \cref{res:utility_representation}. Therefore, $U_{P} (D') > U_P(D^\ast)$.

\subsection{Proof of \texorpdfstring{\cref{res:characterization}}{Theorem \ref{res:characterization}}} \label{sec:proof_characterization}

The proof uses the following lemma. 

\begin{lem}[Effects of delegation parameters]  \label{res:effect_parameters}
	The following hold, within each regime from \cref{res:characterization}.
	\begin{enumerate}[label = \Roman*.]
		\item Hollow: $u_{P,1}$ is strictly decreasing in the gap $y_1 - y_0$ and single-peaked about $y_2 = y_A ( \hat{\th})$; $\D_A$ is strictly increasing in the gap $y_1 - y_0$ and in the cap $y_2$. 
		\item Interval: $u_{P,1}$ is single-peaked about $y_0 = y_A (\hat{\th})$; $\D_A$ is strictly increasing in $y_0$. 
		\item High-point: $u_{P,1}$ is strictly increasing in $y_0$ and strictly quasiconvex in $\bar{y}$; $\D_A$ is strictly convex in $y_0$ and strictly decreasing in $\bar{y}$. 
	\end{enumerate}
\end{lem}

Let $D^\ast$ be an optimal delegation set that is minimal. As shown in the proof of \cref{res:gaps} (\cref{sec:proof_gaps}), $D^\ast$ is not a singleton, so $\max D^\ast > y_A ( \ubar{\th})$. We separate into three cases according to the value of $\max D^\ast$.
\begin{enumerate}[label = \Roman*.]
	\item $y_{A,0} < \max D^\ast \leq y_A (\bar{\th})$. We show that $D^\ast$ is hollow. Let $y_2 = \max D^\ast$. By \cref{res:gaps}, either $D^\ast = [y_A ( \ubar{\th}), y_2]$ or  $D^\ast = [ y_A ( \ubar{\th}), y_0] \cup [y_1, y_2]$ for some $y_0$ and $y_1$ satisfying $y_0 < y_{A,0} < y_1 \leq y_2$ and $\E [ u_A ( y_0, \th)] = \E [ u_A ( y_1, \th)]$.  Either way, we must have $y_2 > y_A ( \hat{\th})$ since, as a function of $y_2$, we know from \cref{res:effect_parameters} that $u_{P,1}$ is single-peaked about $y_A ( \hat{\th})$; $\D_A$ is strictly increasing; and $u_{P,0}$ is constant. 
	
	It remains to prove that $D^\ast$ has a gap. For $r < y_2 - y_{A,0}$, let $D(r) =  [ y_A ( \ubar{\th}), y_0(r)] \cup [y_1(r), y_2]$, where $y_1(r) = y_{A,0} - r$ and $y_0(r)$ is the ex ante conjugate of $y_1(r)$. By the implicit function theorem, $y_1(r)$ is differentiable. It is straightforward to check that $\frac{\de}{\de r} u_{P,1} (D(r))$ is strictly positive at $r  = 0$ because the first-order effects on $\D_A$ and $u_{P,1}$ vanish, and the first-order effect on $u_{P,0}$ is positive (because $y_{A,0} > y_{P,0}$).
	
	\item $y_A ( \ubar{\th}) < \max D^\ast \leq  y_{A,0}$. Let $y_0 = \max D^\ast$. We show that $D^\ast$ is interval. By \cref{res:gaps}, we must have $D^\ast = [ y_A ( \ubar{\th}), y_0]$. It remains to check that $y_0 \neq y_{A,0}$. As a function of the cap $y_0$, both $\D_A$ and $u_{P,1}$ are differentiable at $y_0 = y_{A,0}$, but $u_{P,0}$ is not. The jump from the left to the right derivative is strictly positive since $y_{A,0} > y_{P,0}$.\footnote{The derivative jump is $-(\E[\th] +  a'(y_{A,0})) = \E[ b(\th)]$, which is strictly positive by \ref{it:B}.} Thus, the maximum cannot occur at $y_{A,0}$.
	
	\item $\max D^\ast > y_{A} (\bar{\th})$. Let $\bar{d} = \max D^\ast$. We show that $D^\ast$ is hollow or high-point. By minimality, $\bar{d}$ must be an isolated point of $D^\ast$. Let $y_0 = \max ( D^\ast \setminus \{\bar{d}\})$. By \cref{res:gaps}, we have $D^\ast = [y_A ( \ubar{\th}) \wedge y_0, y_0] \cup \{\bar{d}\}$, where $y_0 < y_{A,0}$ and $\E [ u_A ( y_0, \th)] \geq \E [ u_A (\bar{d}, \th)]$. There are two cases. 
	
	If $\E [ u_A ( y_0, \th)] = \E [ u_A (\bar{d}, \th)]$, then $D^\ast$ is hollow, with $y_1 = y_2 = \bar{d}$. If $\E [ u_A ( y_0, \th)] > \E [ u_A (\bar{d}, \th)]$, then $D^\ast$ is high-point, with $\bar{y} = \bar{d}$. In this case, we prove one further result. Let $\tilde{y}$ be the $\bar{\th}$-conjugate of $\bar{y}$. By single-peakedness, we have $y_0 < \tilde{y} < y_A (\bar{\th})$. We now compare the interval $\tilde{D}= [y_A (\ubar{\th}), y_0] \cup \{\tilde{y}\}$ with $D^\ast$. Clearly, $u_{A,1} (\tilde{D})  > u_{A,1} (D^\ast)$. By the utility representation (\cref{res:utility_representation}), $u_{P,1} (\tilde{D}) > u_{P,1} ( D^\ast)$. Since $D^\ast$ is optimal, we must have $y_{A,0} (\tilde{D}) \neq y_0$, hence $\E [ u_A ( \tilde{y}, \th)] > \E [ u_A ( y_0, \th)]$. 
	
\end{enumerate}

\subsection{Proof of \texorpdfstring{\cref{res:effect_parameters}}{Lemma \ref{res:effect_parameters}}}

\paragraph{I. Hollow} The comparative statics for $u_{P,1}$ are immediate from the utility representation (\cref{res:utility_representation}), and the optimal informed-delegation set (\cref{res:informed_agent_delegation}). Clearly, $\D_A$ is strictly increasing in $y_2$. We show that $\D_A$ is increasing in the gap $y_1 - y_0$. 

It is convenient to parametrize the gap by the left-radius $r$. That is, for each feasible $r$, let $y_0(r) = y_{A,0} - r$, and let $y_1(r)$ be the ex ante conjugate of $y_0(r)$. For $i = 0,1$, let $\th_i(r) = y_A^{-1} (y_i(r))$.\footnote{We can extend the state space so that this is defined.} For $r > 0$, we can differentiate with respect to $r$. We have 
\[
-u_{A,0}'(r) = \E[\th + b(\th)] + a'( y_0(r)) = \th_A + b(\th_A) + a'(y_0(r)).
\]
Apply the implicit function theorem to the ex ante conjugacy condition to get
\begin{equation} \label{eq:IFT}
	y_1'(r) = - \frac{\th_A+ b(\th_A)+ a'(y_0(r))}{\th_A + b(\th_A) +  a' (y_1(r) )}
\end{equation}
We have
\begin{align*}
	u_{A,1}'(r) 
	&= -\int_{\th_0(r)}^{\th_A}  [\th + b(\th) + a'(y_0(r)) ] f(\th) \de \th \\
	&\qquad + y_1'(r) \int_{\th_A}^{\th_1(r)} \Brac{\th + b(\th) +  a'(y_1(r))} f(\th) \de \th \\
	&> -\int_{\th_0(r)}^{\th_A}  [\th_A + b(\th_A)+ a'(y_0(r)) ] f(\th) \de \th \\
	&\qquad + y_1'(r) \int_{\th_A}^{\th_1(r)} \Brac{\th _A+ b(\th_A) + a'(y_1(r))} f(\th) \de \th.
\end{align*}
Simplifying and substituting in \eqref{eq:IFT} gives
\begin{align*}
	u_{A,1}'(r)
	& > - [F(\th_1(r)) - F(\th_0(r))] [ \th_A + b(\th_A) + a'(y_0(r))] \\
	&\geq u_{A,0}'(r),
\end{align*}
as desired. 

\paragraph{II. Interval} Clearly, $u_{P,1}$ is single-peaked about $y_0 = y_A (\hat{\th})$. To see that $\D_A$ is strictly increasing in $y_0$, let $\th_0 = y_A^{-1} (y_0)$. Since $\th + b(\th) + a'(y_0) \geq 0$ if and only if $\th \geq \th_0$, we have
\begin{equation*}
	\begin{aligned}
		u_{P,1}' ( \bar{y}) 
		&= \int_{\th_0 (\bar{y})}^{\bar{\th}} [ \th + b(\th) + a'(\bar{y}] f(\th) \de \th \\
		&> \int_{\ubar{\th}}^{\bar{\th}} [ \th + b(\th) + a'(\bar{y})] f(\th) \de \th \\
		&= u_{A,0}' (\bar{y}).
	\end{aligned}
\end{equation*}

\paragraph{III. High point} From the utility representation, clearly $u_{P,1}$ is strictly increasing in $y_0$. We check that $u_{P,1}$ is strictly quasiconvex in $\bar{y}$. Let $\th^\ast$ denote the  type that is indifferent between $y_0$ and $\bar{y}$. From the utility representation, we have
\begin{align*}
	& u_{P,1}'(\bar{y})  \\
	&= \int_{\th^\ast (\bar{y})}^{\bar{\th}} [ \th + b(\th) + a' (\bar{y})]  [f(\th) + (B(\th) f(\th))' ] \de \th - [ \bar{\th} + b(\bar{\th}) + a' (\bar{y})] B(\bar{\th}) f(\bar{\th}).
\end{align*}
Therefore,
\begin{equation} \label{eq:magic}
	\begin{multlined}
		u_{P,1}'' (\bar{y}) = a '' (\bar{y})  \Brac{ \int_{\th^\ast (\bar{y})}^{\bar{\th}} [f (\th) + (B(\th) f(\th))' ] \de \th - B(\bar{\th}) f(\bar{\th}) } \\
		- \th^{\ast \prime} ( \bar{y})  [ \th^\ast + b(\th^\ast) + a'(\bar{y})] [f(\th^\ast) + (B(\th^\ast) f(\th^\ast))' ].
	\end{multlined}
\end{equation}
The expression on the second line is strictly positive. If $u_{P,1} (\bar{y}) \geq 0$, then the term in brackets in the first line must be nonpositive because
\[
\th + b(\th) + a'(\bar{y}) <\bar{\th} + b(\bar{\th}) + a'(\bar{y}) < 0,
\]
for $\th < \bar{\th}$. Therefore, $u_{P,1}''(\bar{y}) > 0$ whenever $u_{P,1}'(\bar{y}) \geq 0$. Hence, $u_{P,1}$ is strictly quasiconvex. 

Clearly, $\D_A$ is strictly decreasing in $\bar{y}$. We check that $\D_A$ is strictly convex in $y_0$. Let $\th^\ast$ denote the type that is indifferent between $y_0$ and $\bar{y}$. We have
\[
u_{A,0}'(y_0) = \th_A + b(\th_A)  + a' (y_0) = \int_{\ubar{\th}}^{\bar{\th}} [ \th + b(\th) + a'(y_0)]  f(\th) \de \th.
\]
and 
\[
u_{A,1}' (y_0) = \int_{\th_0}^{\th^\ast} [ \th + b(\th) + a'(y_0)] f(\th) \de \th. 
\]
Therefore,
\begin{align*}
	\D_A ' (y_0) 
	&= u_{A,1}'(y_0) - u_{A,0}'(y_0) \\
	&= \int_{\ubar{\th}}^{\th_0} [ \th_0 + b(\th_0) - \th  - b(\th)] f(\th) + \int_{\th^\ast}^{\bar{\th}} [ \th_0 + b(\th_0)  -  \th  - b(\th)] f(\th).
\end{align*}
This expression is strictly increasing in $y_0$ since $\th_0$ and $\th^\ast$ are strictly increasing in $y_0$. 

\subsection{Optimality of high-point delegation} \label{sec:ex_high_point}

Consider the UQC setting. We first outline the argument. Given $\b$ in $(1/4, 1/2)$, we will select suitable thresholds $y_0$ and $\bar{y}$.  Let $D_{\bar{y}} = [\b, 1/2] \cup \{ \bar{y} \}$. We construct a family of cost functions indexed by $\e$. If $\e$ is sufficiently small, so that information acquisition is sufficiently costly, a necessary condition for a delegation set $D$ to be optimal is that  $u_{P,0}(D)$ is sufficiently large. We check that for all $\e$ sufficiently small, the principal strictly prefers the delegation set $D_{\bar{y}}$ to (1) any delegation set $D$ violating the necessary condition (because $u_{P,0}(D)$ is too small); (2) any interval delegation set $D$ satisfying the necessary condition (because $\hat{e}(D)$ is too small); (3) any hollow delegation set satisfying the necessary condition (because $u_{P,1}(D)$ is too small).

Now we turn to the proof proper.  Let $u_0 = -\E[ (1/2 - \th)^2] = -1/12$. For $\bar{y} > 1 + \b$, let $D_{\bar{y}} = [\b, 1/2] \cup \{ \bar{y} \}$.  First, we select the thresholds. 
\begin{lem}[Thresholds] \label{lem:info_gain}
Fix $\b$ in $(1/4,1/2)$. There exist parameters $y_0^\ast$ in $(1/2, 1 - \b)$ and $\bar{y}$ in $(1+ \b, 3/2 + \b)$ such that the following are satisfied:
\begin{align}
		\D_A ( D_{\bar{y}}) &> \D_A ( [\b, y_0^\ast] ), \label{eq:cost_ineq} \\
	 u_{P,1} (D_{\bar{y}}) &> u_0 > u_{P,1} ([\b, y_0^\ast] \cup \{1 + 2 \b - y_0^\ast \} ) \label{eq:cost_u_ineq}.
	\end{align}
\end{lem}

Next, we define the cost function. By \eqref{eq:cost_ineq}, for each $\e$ in $(0,1/2)$ there exists a cost function that induces an effort function $\hat{e} = \hat{e}_\e$ satisfying
\[
	0 < \hat{e} ( \D_A ( [\b, y_0] )) < \e^2 < \e (1 -\e)  < \hat{e} ( D_{\bar{y}}) < \hat{e}(1) < \e.
\]
Since $\hat{e} ( D_{\bar{y}})  > \e (1 - \e)$, it follows from \eqref{eq:cost_u_ineq} that with this cost function,
\begin{equation} \label{eq:util_lower_bound}
\begin{aligned}
U_P(D_{\bar{y}})
& = (1 - \hat{e}(D_{\bar{y}}) ) u_0 + \hat{e}(D_{\bar{y}})  u_{P,1} (D_{\bar{y}}) \\
& > u_0 + \e (u_{P,1} (D_{\bar{y}})  - u_0).
\end{aligned}
\end{equation}
We claim that for all $\e$ sufficiently small, the principal strictly prefers the high-point set $D_{\bar{y}}$ to all hollow and interval delegation sets $D$. We separate into cases. 
\begin{enumerate}
	\item Suppose $u_{P,0} (D) \leq u_{P,0} ([\b, y_0^\ast])$. Since $\D_A(D) \leq 1$, we have $\hat{e}(D) \leq \e$. Since quadratic loss utilities are always negative, 
	\[
		U_P (D) \leq (1 - \hat{e}(D)) u_{P,0} (D) \leq (1 - \e) u_{P,0}([\b, y_0^\ast]).
	\]

	\item Suppose $D$ is an interval delegation set satisfying $u_{P,0} (D) >  u_{P,0} ([\b, y_0^\ast])$. That is, $D =  [\b, y]$ for some $y$ in $(1/2, y_0^\ast)$. By \cref{res:effect_parameters}, $\D_A( [\b, y]) \leq \D_A ([\b, y_0^\ast])$, so $\hat{e} (D) \leq \e^2$. Since $\D_P(D) \leq - u_{P,0}(D) \leq 1$,
	\[
		U_{P} (D)  = u_{P,0}(D) + \hat{e}(D) \D_P (D) \leq u_0 + \e^2.
	\]
	
	\item Suppose $D$ is a hollow delegation set satisfying $u_{P,0} (D) >  u_{P,0} ([\b, y_0^\ast])$. That is, $D = [\b \wedge y_0, y_0] \cup [1 +2 \b - y_0, y_2]$ for some $y_0$ and $y_2$ satisfying $y_0 < 1/2  + \b$ and $y_2 \geq  1 +2 \b - y_0$. Since $u_{P,0} (D) >  u_{P,0} ([\b, y_0])$, we must have $|y_0 - 1/2| < |y_0^\ast - 1/2|$, which implies that $\b < y_0 < y_0^\ast$. By \cref{res:effect_parameters}, the principal's informed-agent payoff $u_{P,1}$ from a hollow delegation set is decreasing in the radius of the gap and single-peaked in $y_2$ about $1 - \b$, which is strictly smaller than $1 + 2 \b - y_0^\ast$ (since $y_0^\ast < 1/2 + \b$ and $\b > 1/4$). Therefore, 
	\begin{equation}
	\begin{aligned}
			u_{P,1} (D)  
			&< u_{P,1} ( [\b, y_0^\ast] \cup [1 + 2\b - y_0^\ast, y_2] ) \\
			&< u_{P,1} ([\b, y_0^\ast] \cup \{1 + 2 \b - y_0^\ast\}).
	\end{aligned}
	\end{equation}
	 From \eqref{eq:cost_u_ineq}, we conclude that $U_P(D) < u_0$. 

\end{enumerate}

Recall that $u_{P,0} ([\b, y_0^\ast]) < u_0$. For all $\e$ sufficiently small,  the lower bound on $U_{P} (D_{\bar{y}})$ in \eqref{eq:util_lower_bound} is strictly greater than the upper bound on $U_P(D)$ in each of the three cases above.

\subsection{Proof of  \texorpdfstring{\cref{lem:info_gain}}{Lemma \ref{lem:info_gain}}}

Fix $\b \in (1/4, 1/2)$. We have 
\[
\lim_{\bar{y}' \uparrow 3/2 + \b} u_{P,1} ( D_{\bar{y}'}) = u_{P,1} ( [\b, 1/2]) > u_0.
\]
Therefore, we may select $\bar{y}$ in $(1+ \b, 3/2 + \b)$ such that 	$u_{P,1} (D_{\bar{y}}) > u_0$. We have 
\begin{equation*}
\begin{aligned}
	\D_A ([\b, 1/2]) &< \D_A ( D_{\bar{y}}), \\
	u_{P,1} ([\b, 1/2] \cup \{ 1/2+ 2\b\})  &= u_0 - (\b/2)(4\b - 1) < u_0.
\end{aligned}
\end{equation*}
Since all the payoffs from the delegation sets $[\b, y_0]$ and $[\b, y_0] \cup \{ 1+ 2\b - y_0\}$ are continuous in $y_0$ over $[1/2, 1/2 + \b)$, we may select $y_0^\ast$ in $(1/2, 1/2 + \b)$ such that
\begin{equation*}
		\D_A ([\b, y_0]) < \D_A ( D_{\bar{y}}), \qquad
		u_{P,1} ([\b, y_0^\ast] \cup \{ 1 + 2\b - y_0^\ast\}) < u_0.
\end{equation*}

\subsection{Proof of \texorpdfstring{\cref{res:delegation_regime}}{Theorem \ref{res:delegation_regime}}}

\begin{lem}[Adding a point] \label{res:adding} Fix decisions $y_0$ and $y_1$, with $y_0 < y_1$. If\/  $\E[ u_A ( y_0, \th)] = \E [ u_A ( y_1, \th)]$, then we have
	\begin{multline*}
		u_{P,1} ( [y_A ( \ubar{\th}) \wedge y_0, y_0] \cup \{y_1\}) \geq u_{P,1} ( [y_A ( \ubar{\th}) \wedge y_0, y_0]) \\
		\iff \E [ \th + b(\th)] \leq \hat{\th} + b(\hat{\th}).
	\end{multline*}
\end{lem}
We use \cref{res:adding} to complete the proof. 

\paragraph{Part 1} Suppose $\E [ \th + b(\th)] \leq \hat{\th} + b (\hat{\th})$. For a contradiction, suppose that $D$ is an optimal delegation set that takes the interval form $[y_A (\ubar{\th}) \wedge y_0, y_0]$. Let $y_1$ be the ex ante conjugate of $y_0$, and set $D' = [y_A (\ubar{\th}) \wedge y_0, y_0] \cup \{y_1\}$. Clearly, $u_{P,0} (D') = u_{P,0} (D)$ and $\D_A (D') > \D_A (D)$. By \cref{res:adding}, $u_{P,1} (D')  > u_{P,1} ( D)$. By optimality, we know $u_{P,1}(D) > u_{P,0} (D)$. Therefore, the principal strictly prefers $D'$ to $D$. 
	
Next, suppose for a contradiction that $D$ is an optimal delegation set that takes the high-point form $[y_A ( \ubar{\th}) \wedge y_0, y_0] \cup \{\bar{y}\}$.  Let $y_1$ be the ex ante conjugate of $y_0$, and set $D' = [ y_A ( \ubar{\th}) \wedge y_0, y_0] \cup \{ y_1\}$. Clearly $u_{P,0}(D')= u_{P,0}(D)$. Since $D$ is optimal, we must have  $u_A (y_1, \bar{\th}) > u_A ( \bar{y}, \bar{\th})$; see the end of the proof of \cref{res:characterization} (\cref{sec:proof_characterization}). Therefore, 
$\D_A( D') > \D_A (D)$. To complete the proof, we show that $u_{P,1}(D') \geq u_{P,1} (D)$. Consider the map $g(y) =  u_{P,1} ([ y_A (\ubar{\th}), y_0] \cup \{y\})$ on the domain $[y_1, \hat{y}]$, where $\hat{y}$ is the $\bar{\th}$-conjugate of $y_0$. In particular, $g(\hat{y}) = u_{P,1} ( [ y_A (\ubar{\th}), y_0])$, so  \cref{res:adding} implies that $g(y_1) \geq g( \hat{y})$. From \cref{res:effect_parameters},  $g$ is  quasiconvex, so $g$ achieves ts maximum at an endpoint, hence at $y_1$. 
Thus, $g(y_1) \geq g(\bar{y})$, as desired. 

\paragraph{Part 2} Suppose $\E [ \th + b(\th)] > \hat{\th} + b (\hat{\th})$. Assume $-\hat{e}''(x)/ \hat{e}'(x) \geq K$ for all positive $x$ (where the constant $K$ will be determined below). Consider a hollow delegation set $D = [y_A ( \ubar{\th}) \wedge y_0, y_0] \cup [y_1, y_2]$, with $u_{P,1}(D) \geq u_{P,1} ([y_A ( \ubar{\th}),   y_{P,0}])$.\footnote{If no such hollow delegation set exist, we are already done.} Let $\tilde{D} = [y_A ( \ubar{\th}),  y_0 \vee y_{P,0}]$. Then $u_{P,0} (\tilde{D}) \geq u_{P,0} (D)$. Therefore, we have $U_{P} (\tilde{D}) > U_P(D)$  if 
\begin{multline*}
	\hat{e}( \D_A (\tilde{D})) [u_{P,1} (\tilde{D}) - u_{P,1} (D)] \\
	> ( \hat{e}( \D_A (D)) - \hat{e}( \D_A (\tilde{D})) [u_{P,1}(D) - u_{P,0}(D)].
\end{multline*}
Since $\hat{e}$ is strictly concave,
\[
	\hat{e}( \D_A (D)) -\hat{e}( \D_A (\tilde{D})) < \hat{e}' ( \D_A(\tilde{D})) [ \D_A(D) - \D_A(\tilde{D})].
\]
Therefore, we have $U_P(\tilde{D}) > U_P (D)$ if 
\begin{equation} \label{eq:strict_ineq}
	\frac{	\hat{e} ( \D_A(\tilde{D})) }{ \hat{e}'(\D_A(\tilde{D})) } \geq \frac{ (\D_A(D) - \D_A(\tilde{D}))(u_{P,1}(D) - u_{P,0}(D)) }{u_{P,1} (\tilde{D}) - u_{P,1} (D)}.
\end{equation}

We claim that there exist positive constants $K_1, K_2, K_3$, depending only on $(F,a,b)$ such that
\begin{align}
	\label{eq:K1} u_{P,1} (\tilde{D}) - u_{P,1}(D) &\geq K_1  (y_2 - y_0), \\
	\label{eq:K2} \D_A ( D) - \D_A (\tilde{D}) &\leq K_2 (y_2 - y_0), \\
	\label{eq:K3} u_{P,1} (D) - u_{P,0} (D) &\leq K_3. 
\end{align}

Now we complete the proof, taking the claim as given. By \cref{res:effect_parameters}, $\D_A(\tilde{D}) \geq \D_A( [y_{A} (\ubar{\th}), y_{P,0}])$. Therefore, \eqref{eq:strict_ineq} holds if 
\begin{equation} \label{eq:inf_ineq}
	\inf_{x} \frac{\hat{e}(x)}{\hat{e}'(x)} > \frac{ K_2 K_3}{K_1}, 
\end{equation}
where the infimum is taken over $x \geq \D_A( [y_{A} (\ubar{\th}), y_{P,0}])$. 

For $x > 0$, let $h(x) = g(x)/g'(x)$. Then 
\[
h'(x) = 1 - \frac{g(x)g''(x)}{(g'(x))^2} = 1 + h(x) \Paren{- \frac{g''(x)}{g'(x)}}. 
\]
Fix a positive number $\a$. If $- g''(x)/g'(x) \geq \a$ for all $x> 0$, then 
\[
h'(x) \geq 1 + \a h(x),
\]
for all $x > 0$. By Gr\"{o}nwall's inequality, 
\[
h(x) \geq  (e^{\a x} - 1)/\a, 
\]
so the infimum in \eqref{eq:inf_ineq} is at least $(1/\a) \exp \Paren{\a \D_A( [y_{A} (\ubar{\th}), y_{P,0}])} - \a^{-1}$, which is larger than the right side of \eqref{eq:inf_ineq}, provided that $\a$ is chosen sufficiently large.

\paragraph{Part 2---Proof of claim} Define $\th_A$ by $y_{A}'(\th_A) = \th_A + b(\th_A)$. From the proof of \cref{res:informed_agent_delegation} (\cref{sec:proof_informed_agent_delegation}), we have
\begin{equation} \label{eq:M}
\begin{aligned}
	u_{P,1} ([y_A ( \ubar{\th}) \wedge y_0,  y_0] \cup \{y_1\} ) - u_{P,1} ( D) 
	&=
	u_{P,1} ( [y_A(\ubar{\th}), y_1] ) - u_{P,1} ( [y_A(\ubar{\th}), y_2] )  \\
	&\geq 
	M (y_2 - y_1),
\end{aligned}
\end{equation}
where $M$ is the minimum, over $\hat{\th}$  in $[\th_A, \bar{\th}]$, of the negative of the derivative in \eqref{eq:deriv}. This minimum is strictly positive by continuity. From the computation in the proof of \cref{res:adding} (\cref{sec:proof_adding}), 
\begin{equation} \label{eq:M'}
\begin{aligned}
	&u_{P,1} ([y_A ( \ubar{\th}) \wedge y_0,  y_0] - u_{P,1} ([y_A ( \ubar{\th}) \wedge y_0,  y_0] \cup \{y_1\} )  = M' (y_1 - y_0),
\end{aligned}
\end{equation}
where 
\[
	M' = - \bigl(  \th_A +  b(\th_A) - \E [ \th | \th \geq \th_A] \bigr)  (1 - F(\th_A)). 
\]
Let $K_1 = M \wedge M'$. By \cref{res:effect_parameters}, we have $u_{P,1} ( \tilde{D}) - u_{P,1} ([y_A ( \ubar{\th}) \wedge y_0,  y_0] > 0$. Summing this inequality with \eqref{eq:M'} and \eqref{eq:M} gives \eqref{eq:K1}. 

From the computation in the proof of \cref{res:adding} (\cref{sec:proof_adding}), 
\begin{equation*}
\begin{aligned}
	&u_{A,1} ( D) - u_{A,1} (\tilde{D})  \\
	&= \E [ u_{A} ( y_A (D, \th), \th) - u_A ( y_0, \th) | \th \geq \th_A] (1 - F(\th_A))\\
	&\leq  \E [ (\th + b(\th) - \ubar{\th} - b(\ubar{\th})) ( y_{A} (D, \th) - y_0) | \th \geq \th_A] (1 - F(\th_A))\\
	&\leq   \Paren{ \E [ \th + b(\th) | \th \geq \th_A] - \ubar{\th} - b(\ubar{\th}) } (y_2 - y_0) (1 - F(\th_A)).
\end{aligned}
\end{equation*}
This gives \eqref{eq:K2}. Finally, \eqref{eq:K3} is immediate since the inequality $u_{P,1}(D) > u_{P,1} ([y_A ( \ubar{\th}),   y_{P,0}])$ ensures that  $y_0 $and $y_1$ are bounded.

\paragraph{Part 3} If $u_{P,1} ( \{ y_{P,0}\}) \geq u_{P,1} ([ y_A ( \ubar{\th}), y_{A,0}])$, then $y_{A,0} > y_{A} (\hat{\th})$. Therefore, 
  for any hollow delegation set $[y_A(\ubar{\th}) \wedge y_0, y_0] \cup [y_1, y_2]$, we have
\begin{align*}
	u_{P,1} ( [y_A(\ubar{\th}) \wedge y_0, y_0] \cup [y_1, y_2])  
	&\leq 
	u_{P,1} ( [y_A(\ubar{\th}), y_2]) \\
	&<
	u_{P,1} ( [y_A(\ubar{\th}), y_{A,0}]) \\
	&\leq 
	u_{P,1} ( \{ y_{P,0}\}).
\end{align*}
Therefore, the principal strictly prefers $\{ y_{P,0}\}$ to every hollow delegation set.

\subsection{Proof of \texorpdfstring{\cref{res:adding}}{Lemma \ref{res:adding}}} \label{sec:proof_adding}

Define $\th_A$ by $\th_A + b(\th_A) = \E [ \th + b(\th)]$. Fix decisions $y_0$ and $y_1$ with $y_0 < y_1$. If $\E [ u_A ( y_0, \th) ] = \E [ u_A (y_1, \th)]$, then $u_A ( y_0, \th_A) = u_A (y_1, \th_A)$, so 
\[
a(y_1) - a(y_0) = - (\th_A + b(\th_A))(y_1 - y_0).
\]
For any state $\th$, we have 
\begin{align*}
	u_P (y_1, \th) - u_P (y_0, \th)   
	&= a(y_1) - a(y_0) +  \th ( y_1 - y_0)  \\
	&= (- \th_A - b(\th_A)+ \th ) (y_1 - y_0).
\end{align*}
Therefore, 
\begin{equation*}
	\begin{aligned}
		&u_{P,1} ( [y_A ( \ubar{\th}) \wedge y_0, y_0] \cup \{y_1\}) - u_{P,1} ( [y_A ( \ubar{\th}) \wedge y_0, y_0]) \\
		&= \E [  u_P(y_1, \th) - u_P(y_0, \th)|\th \geq \th_A] (1 - F(\th_A)) \\
		&= \bigl( -\th_A - b(\th_A)  + \E[\th | \th \geq \th_A ] \bigr) (y_1 - y_0) (1 - F(\th_A)).
	\end{aligned}
\end{equation*}
The last line is nonnegative if and only if $\E [ \th | \th \geq \th_A] \geq \th_A + b(\th_A)$. By the proof of \cref{res:informed_agent_delegation}, this inequality holds if and only if $\th_A \leq \hat{\th}$, or equivalently, $\E [ \th + b(\th)] = \th_A + b(\th_A) \leq \hat{\th} + b(\hat{\th})$. 

\subsection{Proof of \texorpdfstring{\cref{res:CS}}{Theorem \ref{res:CS}}}

Following the proof of \citet[Proposition 3, p.~1181]{szalay2005economics}, it can be shown that if \eqref{eq:Szalay_cost} holds, then with an unbiased agent, there exists an optimal delegation set $D^\ast = [0 \wedge (1/2 - r),1/2 -r] \cup [1/2 + r, 1 \vee (1/2 + r)]$ for some strictly positive radius $r$.\footnote{Our conclusion is slightly weaker than that in \citet[Proposition 3, p.~1181]{szalay2005economics} because we do not restrict to delegation sets within $[y_A(\ubar{\th}), y_A( \bar{\th})]$. If the state is uniformly distributed on $[0,1]$, then $u_{P,1} ( \{0,1\}) < u_{P,1} (\{ 1/2\})$, so the optimal radius must be strictly smaller than $1/2$.} For all $\b \geq 0$, let $D( \b) = \b + D^\ast$.  If the principal offers the delegation set $D(\b)$ to an agent with bias $\b$, then the agent's return from effort is independent of $\b$.  Denote this effort level by $e^\ast$. Let $V(\b)$ denote the value of the principal's delegation problem when facing an agent with bias function $\b$. Write  $U_P (\cdot ; \b)$, $u_{P,0} (\cdot; \b)$, and $u_{P,1} ( \cdot ; \b)$ to  denote payoffs that depend on the agent's bias $\b$. We have
\begin{equation*}
	\begin{aligned}
		&V(\b)  - V(0)\\
		&\geq U_P (D(\b); \b)  - U_P (D^\ast, 0) \\
		&= (1 - e^\ast) [u_{P,0} (D(\b); \b) -u_{P,0} (D^\ast)]  + e^\ast [u_{P,1} ( D(\b); \b)  - u_{P,1} (D^\ast, 0) ] \\
		&= (1 - e^\ast)   \Brac{ - (\b - r)^2 + r^2} + e^\ast \E \Brac{ - ( y_A ( D^\ast, \th) + \b - \th)^2  + (y_A ( D^\ast, \th) - \th)^2}.
	\end{aligned}
\end{equation*}
In the last term, expand the square to get
\[
	 (y_A ( D^\ast, \th) + \b - \th)^2 = \b^2 + \b (y_A ( D^\ast, \th) - \th) +  (y_A ( D^\ast, \th) - \th)^2.
\]
On the right side, the expectation of the middle term vanishes, provided that
\[
	\E \Set{ (\th -\E\th) [ \E\th \leq \th \leq \E \th + r]}
	=
	\E \Set{( \E\th - \th) [\E \th - r \leq \th \leq ]},
\]
which is guaranteed by the symmetry assumption.  We conclude that 
\begin{equation*}
\begin{aligned}
	V(\b) - V(0) 
	&\geq (1 - e^\ast)   (- (\b - r)^2 + r^2) - e^\ast \b^2 \\
	&= \b \Brac{ 2 r (1 -e^\ast)  - \b }.
\end{aligned}
\end{equation*}
Let $\bar{\b}  = 2r (1 - e^\ast)$. We conclude that $V(\b) > V(0)$ for all $\b$ in $(0, \bar{\b})$.

\newpage
\bibliographystyle{ecta}
\bibliography{literature}

\end{document}